\documentclass[twocolumn,superscriptaddress,altaffilletter,prd]{revtex4-1}
\usepackage[utf8]{inputenc}
\usepackage{import}
\usepackage{siunitx}
\usepackage{amssymb}
\usepackage{amsmath}
\usepackage{multirow}
\usepackage[shortlabels]{enumitem}
\usepackage{hyperref}
\usepackage[capitalise]{cleveref}

\usepackage{hepparticles} 
\usepackage[italic]{heppennames}
\usepackage{graphicx}
\usepackage[caption=false]{subfig}
\usepackage{boxhandler}
\usepackage[dvipsnames]{xcolor}

\def\geantFour  {\mbox{\scshape Geant4}\xspace}

\usepackage[draft]{pgf}
\usepackage{tikz}

\begin{document}

\title{Simulations of Events for the LUX-ZEPLIN (LZ) Dark Matter Experiment}

\author{D.S.~Akerib}
\affiliation{SLAC National Accelerator Laboratory, Menlo Park, CA 94025-7015, USA}
\affiliation{Kavli Institute for Particle Astrophysics and Cosmology, Stanford University, Stanford, CA  94305-4085 USA}

\author{C.W.~Akerlof}
\affiliation{University of Michigan, Randall Laboratory of Physics, Ann Arbor, MI 48109-1040, USA}

\author{A.~Alqahtani}
\affiliation{Brown University, Department of Physics, Providence, RI 02912-9037, USA}

\author{S.K.~Alsum}
\affiliation{University of Wisconsin-Madison, Department of Physics, Madison, WI 53706-1390, USA}

\author{T.J.~Anderson}
\affiliation{SLAC National Accelerator Laboratory, Menlo Park, CA 94025-7015, USA}
\affiliation{Kavli Institute for Particle Astrophysics and Cosmology, Stanford University, Stanford, CA  94305-4085 USA}

\author{N.~Angelides}
\affiliation{University College London (UCL), Department of Physics and Astronomy, London WC1E 6BT, UK}

\author{H.M.~Ara\'{u}jo}
\affiliation{Imperial College London, Physics Department, Blackett Laboratory, London SW7 2AZ, UK}

\author{J.E.~Armstrong}
\affiliation{University of Maryland, Department of Physics, College Park, MD 20742-4111, USA}

\author{M.~Arthurs}
\affiliation{University of Michigan, Randall Laboratory of Physics, Ann Arbor, MI 48109-1040, USA}

\author{X.~Bai}
\affiliation{South Dakota School of Mines and Technology, Rapid City, SD 57701-3901, USA}

% ...

\collaboration{The LUX-ZEPLIN (LZ) Collaboration}
\date{\today}

\begin{abstract}
The LUX-ZEPLIN dark matter search aims to achieve a sensitivity to the WIMP-nucleon spin-independent cross-section down to (1--2)$\times10^{-12}$\,pb at a WIMP mass of 40 GeV/$c^2$. This paper describes the simulations framework that, along with radioactivity measurements, was used to support this projection, and also to provide mock data for validating reconstruction and analysis software. Of particular note are the event generators, which allow us to model the background radiation, and the detector response physics used in the production of raw signals, which can be converted into digitized waveforms similar to data from the operational detector. Inclusion of the detector response allows us to process simulated data using the same analysis routines as developed to process the experimental data.
\end{abstract}

\maketitle



\section{Introduction}
\label{sec:Intro}

The LUX-ZEPLIN (LZ) dark matter direct detection experiment \cite{akerib:2015cja,LZ-TDR,LZ-sensitivity,LZ-detector} will use a dual-phase, 7 tonne liquid-gas xenon time projection chamber (TPC). The goals of LZ encompass primarily the pursuit of Weakly Interacting Massive Particles (WIMPs) through spin-independent and spin-dependent interactions with target xenon nuclei, but also axions and axion-like particles, searches for additional dark matter candidates using effective field theory (EFT), and neutrino-less double-beta decay. The signal for one search is sometimes the background of another. However, common backgrounds arise in the form of intrinsic radioactivity from detector components (including surface contamination), noble radioisotope impurities present in the xenon, and environmental and cosmogenic radiation.

To achieve a high sensitivity to rare events, any background must be heavily suppressed. In this respect, LZ will benefit from: 
\begin{enumerate}[a)]
\itemsep0em 
\item the underground location of the experiment (4850~ft below the surface at the Sanford Underground Research Facility (SURF), with a mean slant depth of about 4.5~km~water~equivalent);
\item shielding provided by a tank of water surrounding the detector, of height 591~cm and radius 381~cm; 
\item an active veto system including an instrumented outer layer of about 2 tonnes of liquid xenon (the LXe skin), around 17 tonnes of gadolinium-doped, organic liquid scintillator (GdLS) in the outer detector (OD) tanks, and water in the tank mentioned in b); 
\item the self-shielding properties of the xenon and accurate position reconstruction, which allow for fiducialization of the LXe volume;
\item identification and rejection of events with multiple scatters;
\item accurate energy reconstruction;
\item discrimination between nuclear recoils, as expected from WIMPs, and electron recoils from gamma interactions, beta-decays and some types of signal (e.g. axions). 
\end{enumerate}

For the main aim of a WIMP search, the sensitivity to the WIMP-nucleon spin-independent cross-section is predicted to reach $1.4\times10^{-12}$~pb at 40~GeV/$c^{2}$ WIMP mass \cite{LZ-sensitivity} due in large part to these features. This represents the smallest cross-section that will be probed in a direct detection experiment, based on the projected sensitivities of other experiments in operation or construction.      

Simulations play a key role in estimating our background rejection efficiency, and predicting the residual background based on radioactivity measurements and known external sources (e.g. solar neutrinos or cosmic-ray muons). At their core, they must be capable of capturing any part of the expected particle flux, be it signal or background, and the response it produces in the detector media. For LZ, these media consist of the liquid and gaseous xenon in the TPC and skin, and the GdLS employed in the outer detector. The skin and OD together comprise an effective veto system, the former ideal for tagging scattered $\gamma$-rays, the latter highly efficient as a neutron veto because of the very high thermal neutron capture cross-section of gadolinium and the high total energy released in such captures. The veto system is complemented by the instrumented water, which is able to catch Cherenkov radiation from muons or muon-induced cascades if they miss other active media of the detector.

Within the TPC, for low-energy (up to a few MeV) localized events, a recoil of either a xenon nucleus or an electron is detected with a prompt scintillation signal (S1) followed by a delayed charge signal (S2). The S2 is formed by drifted ionization electrons extracted with high efficiency into the gas phase, where they cause electroluminescence. The ratio of these two signals allows for discrimination between nuclear and electron recoils, whilst the time between them provides the depth of the interaction. The xenon target is viewed by 253 photomultiplier tubes (PMTs) at the top of the cryostat, and 241 PMTs on the bottom. The hit pattern of light on these PMTs, particularly the localization of the S2 signal on the top array, allows for effective 3D position reconstruction of the interaction vertices. The main design features of the LZ detector are described in greater detail in \cite{akerib:2015cja,LZ-TDR,LZ-sensitivity,LZ-detector}.

In this paper we focus on the method of simulating signals for different types of particles. Such simulations are based on an in-house software package, BACCARAT (Section~\ref{sec:LZSim}), that tracks particles using \geantFour \cite{Geant4}. Various features have been added to BACCARAT to better model the xenon and GdLS response (Section~\ref{sec:LightAndCharge}). A second package, the DER (Section~\ref{sec:DER}) exists to reproduce the signal processing done on the resulting PMT hits. Primary particles are specified using generators, be they for backgrounds, calibration or physics sources. Sections~\ref{sec:Backgrounds},\ref{sec:Calibrations},\ref{sec:Neutrinos}) discuss the simulations and studies done with a number of these generators, which have been incorporated into the backgrounds model and sensitivity analyses.


\section{Simulations Framework}
\label{sec:SimulationsFramework}
\subsection{Overview}
\begin{figure}[htb]
\begin{center}
\includegraphics[width=8.5cm]{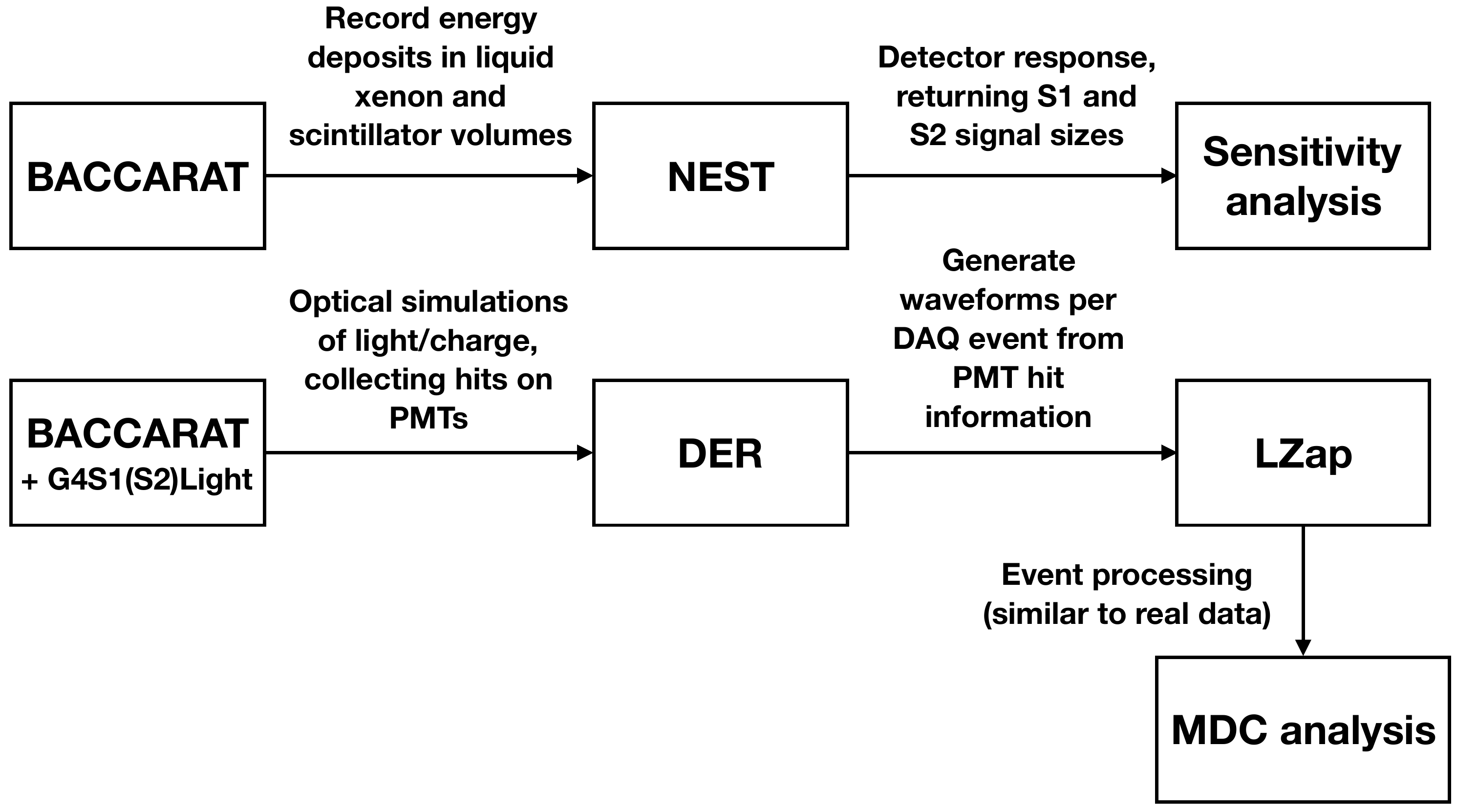}
\caption {Processing chain for simulations used to generate data for the background model (sensitivity analyses) and Mock Data Challenges (MDCs).}
\label{fig:simsChain}
\end{center}
\end{figure}

Monte Carlo modeling of events for LZ serves several purposes: the assessment of design features of the detector through, for example, efforts to maximize the light collection efficiency; the calculation of the rate of background events in LZ with input from radioactivity measurements; the prediction of the sensitivity of the experiment to various rare event searches based on the background rate and Profile Likelihood Ratio analysis (PLR); the simulation of the whole event processing chain for future reconstruction validation and tests. All of these simulations begin with BACCARAT, which tracks particles using \geantFour and identifies their interaction points in the detector (Section~\ref{sec:LZSim}). From there, two separate chains exist for consuming this information (Figure~\ref{fig:simsChain}).

The first chain records the energy deposits in the detector and passes them to NEST (Section~\ref{sec:NEST}) to generate S1 and S2 signals. This enables large statistics datasets to be generated, which can be bulk analyzed to assess, for example, background rates and inform sensitivity estimates. The downside is that these results rely on detector-averaged quantities and do not contain information on the times of interactions or of photon hits on PMTs.

The second chain enables a full simulation of the VUV photons and ionization electrons that are produced during xenon interactions, as well as the scintillation light generated in the OD. The majority of the physics involved is not well-modeled in \geantFour, and thus external models have been added into BACCARAT to capture these phenomena (Section~\ref{sec:LightAndCharge}). NEST is then referenced in BACCARAT solely to calculate the raw photons and electrons emerging from an interaction.

The PMT hits recorded in the BACCARAT output are then translated into waveforms by the DER (Section~\ref{sec:DER}), which simulates the PMT response and transforms the signal as if it had been transferred through the various stages of the read-out electronics. The resulting data are organised in the DAQ event format, which means they can be passed through the event processing framework and analyzed much like real data. Whilst this chain is more computationally intensive, it allows for more realistic, event-by-event analysis.

Simulations with the second chain are carried out for Mock Data Challenges (MDCs), in which analyzers are presented with data designed to mimic what is expected in commissioning and science run periods. The sources simulated are the major radioactive components, with their activities set to approximate the expected background event rate, calibration sources, and potentially WIMP and non-WIMP physics signals (Sections~\ref{sec:Backgrounds}--\ref{sec:Neutrinos}). Detector parameters, such as the optical reflectivities of different surfaces and the attenuation length of drifting electrons, are also defined and may be time dependent. The ultimate goal of the MDCs is to ensure physics readiness when data-taking commences through the development of data analysis methods and reconstruction tools.

\subsection{Particle Generation and Tracking: BACCARAT}
\label{sec:LZSim}
\begin{figure}[htb]
\begin{center}
\includegraphics[width=8cm]{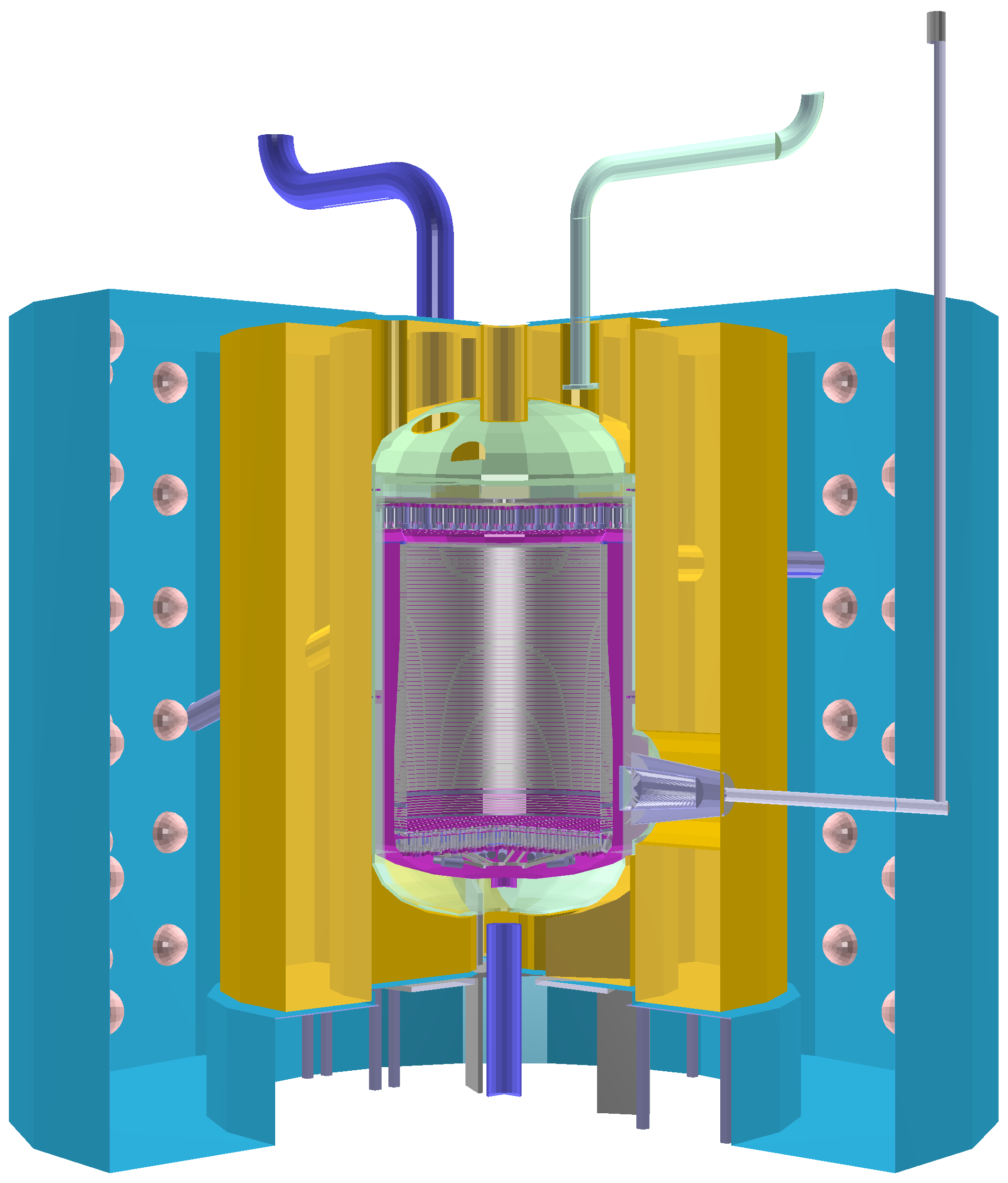}
\caption {Visualization of the LZ detector in \geantFour-based BACCARAT. The TPC (magenta) within the cryostat (light green) is surrounded by the GdLS outer detector tanks (yellow) that are immersed in a water tank. The inner part of the water tank within the PMT structure is shown in blue.}
\label{fig:lz-vis}
\end{center}
\end{figure}

The LZ code for simulating particles and their interactions, BACCARAT (Basically, A Component-Centric Analog Response to AnyThing), builds upon that developed for LUX \cite{luxsim:2012}, which sought to provide a more useful interface to \geantFour for low-background experiments. Central to this interface is the shift of focus towards individual volumes in the geometry (components). A C++ detector 
component object was implemented that inherits from the underlying \geantFour structure with the goal of making the code familiar to users of this software, but also providing additional functionality. This functionality includes macro-level control to set a component as a three-dimensional source of radioactivity for modeling impurities in the materials or surface contaminants, as well as the ability to record varying levels of information about what occurs in each component, such as the total energy deposited. Figure~\ref{fig:lz-vis} illustrates the geometry of the LZ detector components defined with BACCARAT.

A suite of custom-built generators, based on the standard framework and generators available within \geantFour, is used to produce the various types of particles that might interact in the detector, including those from radiogenic decay and cosmic ray events. They allow for the simultaneous and time-delayed emission of multiple primary particles from a given source. Several sources may be loaded onto the same or different components, and with arbitrary activities, to replicate the total expected particle flux in a single simulation run. The primary particles are pre-determined in time and location and chronologically ordered at the beginning of the simulation, allowing for realistic position and time-based analysis of the output data stream. A similar approach is described in Ref.~\cite{kareem2011}. The resultant particles can go on to interact in the detector media and form a single event. The beginning time of each event is recorded to allow for higher-level analysis, such as event pile-up. 

The \geantFour toolkit contains pre-defined physics lists that provide options for modeling various processes, intended to align with a specific application. The toolkit also contains the functionality to allow user-defined processes to be integrated into the physics of the simulation. The prominent modules that are deployed in BACCARAT are:
\begin{enumerate}
\item \textbf{G4EMLivermorePhysics}, which covers electromagnetic interactions using Livermore models for gamma and electron cross-sections \cite{liv1} \cite{liv2}, extending the validity of the physics down to 10\,eV. This has a particular focus on low energy processes, such as Rayleigh and Compton scattering, bremsstrahlung and the photoelectric effect; 
\item \textbf{G4HadronPhysicsQGSP\_BIC\_HP\_Gd}, which uses the Binary Cascade (BIC) intra-nuclear model \cite{Folger2004} for certain lower energy inelastic interactions, and adds several community modifications that better model nuclear processes on Gd: this includes the DICEBOX neutron capture model described in Section~\ref{sec:DB};
\item \textbf{G4S1Light \& G4S2Light}, which have been developed by LZ collaborators to integrate NEST physics into BACCARAT (Section~\ref{sec:NEST}), and which govern the generation of light and charge quanta in the xenon.
\end{enumerate}
These are in addition to a set of standard reference physics lists that determine, amongst other things, the production of light via scintillation and Cherenkov processes in non-xenon materials; the at rest and in-flight decay of radioactive nuclei via $\alpha$, $\beta^{\pm}$, $\gamma$ emission or electron capture; the emission of electrons and X-rays due to the relaxation of excited atomic states; the hadronic interactions of photons, electrons and positrons. Both the generators and the physics lists have been substantially developed since the advent of BACCARAT in an effort to better construct events that might be seen in LZ. Examples of custom-built generators include muon events, wall events, coincident neutrons and gammas from ($\alpha,n$) reactions, a number of calibration sources etc. Some of them will be described in more detail below.

The code base continues to be expanded and maintained via a Git version control system, and tested against newer versions of \geantFour, with BACCARAT verified against 10.3 at the time of writing. Given the goal of increased realism of the simulations through the incorporation of the latest physics models and understanding, and to provide added functionality and streamline the process for users, we periodically evaluate additional releases of \geantFour for production running of official simulations. In addition, background pathologies which cannot be simulated by a simple event generator are also considered, such as single-electron backgrounds (Section \ref{sec:EB}).

\subsection{Simulating Light and Charge Response}
\label{sec:LightAndCharge}
\subsubsection{Noble Element Simulation Technique (NEST)}
\label{sec:NEST}

The production of both VUV scintillation photons and thermal ionization electrons is modeled with the NEST~\cite{NEST1,NEST2,mock:2013ila,Lenardo:2014cva} formalism, a semi-empirical collection of models based on past and present detectors' calibration and science data sets, as well as on the specific measurements from purpose-built instruments. NEST is frequently updated based on the most up-to-date published data, with new features/tools also added regularly; it is both postdictive and predictive~\cite{akerib:2013tjd,Akerib:2015rjg,akerib:2017hph}.

NEST simulates the excitation, ionization, electron recapture (``recombination''), and electron electroluminescence processes in liquid or in gaseous xenon, as a function of particle and interaction type, energy, ionization density, stopping power, electric field~\cite{Akerib:2016vxi,Akerib:2017btb}, and fluid density, via the temperature and pressure. NEST also models S1 and S2 pulse shape profiles versus time, based on known light and charge yields~\cite{Akerib:2017vbi,Akerib:2018psd}. The contemporary NEST version, v2.0, uses simple sigmoidal-class functions to model yields as functions of energy inspired by the Doke modification to Birks' law~\cite{Doke} at high energies and the Thomas-Imel box recombination model~\cite{TIB} at low energies, with coefficients as functions of field. The quantum generation is vetted against data from 1 to 1000~keV for betas, 0.1--5000~keV for gammas, 0.5--300~keV for neutron-induced xenon recoils, and fields from 0 to more than 10000~V/cm. 

Multiple steps determine how simulated quanta are translated into the observables of primary (S1) and secondary (S2) scintillation light. The exciton-to-ion ratio first determines how many excitons, leading to S1, and ion-electron pairs are initially generated by an energy deposit. Low-energy electrons recombine to produce additional excitons, which add to the S1 signal after de-excitation. A third possible channel is also taken into account: energy loss into heat, not visible in a TPC. This is represented by a simple power law that closely approximates the Lindhard factor~\cite{lindhard:1961zz,lindhard:1963}. 

The electrons which instead escape are drifted to the liquid-gas interface. NEST can handle transportation under a parameterized electric field applying diffusion and a finite mean free path, which stems from the concentration of impurities; a separate set of drift simulations was devised to examine this with a variable electric field model, as described in Section~\ref{sec:charge}. Surviving electrons, successfully extracted into the gas, will produce electroluminescence (S2), with the amount of light being a function of electric field and pressure in the gas phase. The probability of extraction is quantified in NEST via an extraction efficiency, although there are theorized mechanisms that can delay the emission and thus affect the S2 signal (Section~\ref{sec:EB}).

\begin{figure}[!h]
\begin{center}
\includegraphics[ width=8.5 cm, trim = 40 0 45 -10]{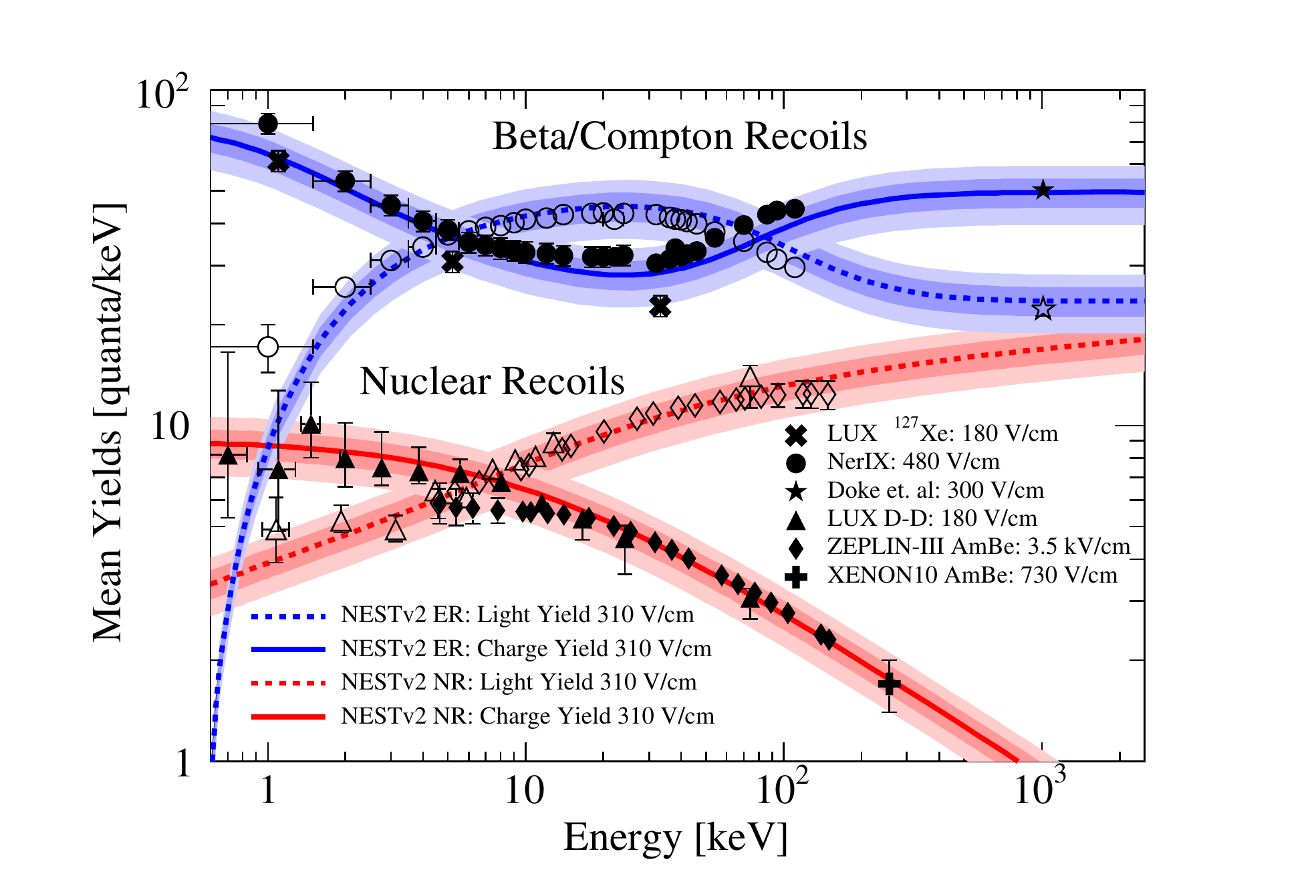}
\includegraphics[ width=8.5 cm, trim = 40 0 45 -10]{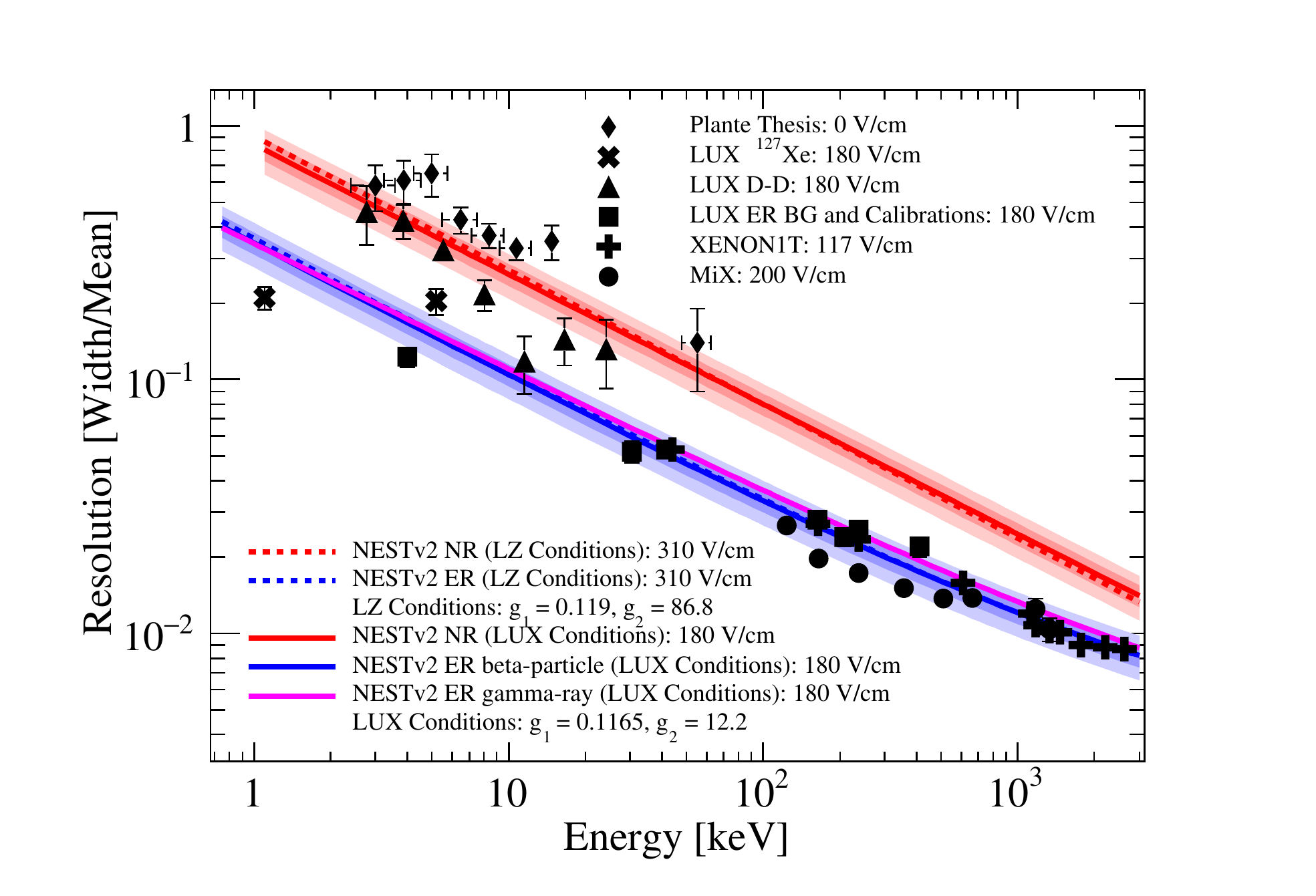}
\caption{Top: the derived average ionization charge (full markers) and scintillation light (open markers) yields from representative world calibration and background data sets. NEST v2.0 detector-independent yields for NR (red) and ER (blue) are shown for the baseline LZ drift field of 310 V/cm, with bands indicating the estimated 1 and 2 $\sigma$ systematic uncertainty. Bottom: the energy resolution from NEST for NR (red), $\beta$-only ER (blue) and $\gamma$/X-ray (green) under LZ (dashed) and LUX (solid) conditions, again compared to experimental data (solid markers). Resolutions in NEST follow a $1/\sqrt(E)$, with the cut-off at low energy corresponding to detector threshold, and the S1 able to remain above threshold down to lower energies.}
\label{fig:lzNESTfig}
\end{center}
\end{figure}

NEST can be run both within BACCARAT, or as a standalone executable (fastNEST). To generate the final signals, raw photons and electrons can be passed back to BACCARAT for transport (Section~\ref{sec:charge}) and ray-tracing propagation to the PMTs (Section~\ref{sec:optical}), or translated into S1 and S2 signals using detection efficiencies derived from BACCARAT in fastNEST. fastNEST bypasses the processing chain of full mock data, and thus enables rapid generation of high-statistics simulated data for the signal and background models~\cite{akerib:2014rda,Akerib:2016vxi,Akerib:2017uem} entering into sensitivity, limit, or discovery calculations. 

In summary, NEST creates non-analytical NR and ER yield probability distribution functions with  high-precision S1/S2 means and energy resolution~\cite{Akerib:2016qlr}, as illustrated in Figure~\ref{fig:lzNESTfig}, which compares NEST with selected pre-LZ empirical data.

\subsubsection{Charge Transport} 
\label{sec:charge}
The positional distribution of the S2 light depends upon the trajectories of the ionization electrons that cause the electroluminescence. These electrons have too low a kinetic energy for their drift to be handled efficiently by \geantFour transportation processes. Therefore a bespoke simulation of the electron drift was devised, whose results are used in BACCARAT to predict the drift time and radial locations of the electrons on the liquid-gas surface.
 
A radially symmetric QuickField model \cite{quickfield} of the TPC electric field was used as input for the drift calculation. Different electric field configurations can be considered by modifying the grid potentials. The electric field strength at any point along the electron trajectory can be found by sampling the QuickField map. The electron drift velocity can be found from the field strength using a function determined by the EXO collaboration \cite{EXO}. This allowed for the evaluation of the instantaneous drift velocity of an electron at any given location in the LXe. For the particular field strengths simulated, this saw an average drift velocity of 1.6 mm/$\mu$s in the bulk. Thus the average drift path of electrons from an initial position (sampled at $1$~mm intervals throughout the LXe), to a final position at the phase boundary was calculated. The velocity of the virtual electron was updated every 1~$\mu$s in order to closely approximate the true mean velocity and path of an electron traversing the LXe.

The simulations resulted in a look-up table giving the electron drift time and final radial location for given positions of origin in the LXe. This map was interpolated in BACCARAT to fold the field into the simulated S2 signals. To validate this procedure, the data was unfolded using the same step-wise method to reconstruct the original positions of each interaction in the TPC.

This initial calculation disregarded the stochastic process of electron diffusion which was handled within NEST. Diffusion during drift will perturb the trajectories of the electrons: non-uniformities in the field lead to a non-Gaussian final position distribution of electrons originating from the same interaction point.

\subsubsection{Electron Pathologies}
\label{sec:EB}
Delayed, spontaneous and induced emissions of electrons can produce pathological features in the data. Tails of single electrons (SEs) following an S2 signal can alter the reconstructed parameters of an event, leak into successive events producing pile-up or even be triggered on as standalone events. They can also create a considerable background to S2-only analyses focused on low-mass WIMPs. Several sources of SEs have been identified and introduced in the simulation. 

Firstly, VUV light from Xe scintillation or electroluminescence is energetic enough to extract electrons from the grid wires via the photoelectric effect. The hit pattern of S2 photons on the wires was modeled and introduced into the simulation. The average number of S2 photons per electron and an estimate of the grid wire QE \cite{wires} are used in a Poisson random number generator. The number of produced SEs drawn are then placed at the appropriate locations according to the hit distribution. 
 
VUV photons may also ionize impurities dispersed in the liquid xenon. An estimate from LUX of the bulk photoionization yield of $5\times10^{-5}$ SEs per detected photon \cite{luxmeasurements} was scaled by the LUX photon detection efficiency and the ratio of expected impurity levels in LZ with respect to that inferred in LUX. This gives an estimated bulk photoionization probability of $7\times10^{-5}$ per photon in LZ, again used in a Poisson distribution to determine the number of SEs generated from S2 light. These were placed homogeneously throughout the LXe following the original S2 signal.

Next, two exponential tails following an S2 signal have recently been characterized \cite{Sorensen:2017}. The fast component, with a time-constant of $\sim$30~$\mu$s, is attributed to electrons trapped at the liquid-gas interface due to an extraction efficiency smaller than unity. A random $3\%$ of ionization electrons pick up a time delay to follow this exponential. The slow component has been simulated with a time constant of $\sim$20~ms and is applied to $1\%$ of ionization electrons. Several explanations for this effect can be found in ~\cite{Sorensen:2017}.

Other effects that could cause electron backgrounds exist but have not yet been implemented due to their smaller contributions, for example: electrons from field emission from the grid wires and photoionization of impurities due to PTFE fluorescence. The total rate of SEs from these sources is expected to be $\mathcal{O}(1$ Hz), based on observations from ZEPLIN-III \cite{SEZeplinIII}.

\subsubsection{GdLS Optical Physics}
Optical simulation of the outer detector is non-trivial due to the complex geometry of the 10 segmented acrylic tanks and scintillation properties of the GdLS. The scintillation light, which ranges in wavelength from 350~nm to 550~nm, can travel through the acrylic and be collected by 120 PMTs situated on a tyvek curtain 115~cm from the outer radius of the tanks. Data from a small prototype detector, the LS Screener~\cite{Haselschwardt:2018}, which was calibrated with various $\alpha$-decays, $\beta$-decays and $\gamma$-rays, was used to fine-tune a modified version of the \geantFour G4Scintillation code. 

Firstly, generation of photons from energy deposits in the scintillator is implemented using a modified Birk's law formula~\cite{vonKrosigk:2015aaa}. The mean number of photons emitted along the particle track, $\textrm{d}L/\textrm{d}x$, is given by:
\vspace{0.2cm}
\begin{equation}
\label{eq:Birks}
\frac{\textrm{d}L}{\textrm{d}x} = Y \frac{ \frac{\textrm{d}E}{\textrm{d}x} }{ 1 + kB \left(\frac{1}{\rho} \frac{\textrm{d}E}{\textrm{d}x}\right) + C \left(\frac{1}{\rho} \frac{\textrm{d}E}{\textrm{d}x}\right)^{2} }
\end{equation}
where $\rho$ is the density of the GdLS, 0.86~g/cm$^3$, $\textrm{d}E/\textrm{d}x$ is the energy loss per unit path length, $Y$ is the scintillator light yield (i.e. photons/MeV), $kB$ and $C$ are the first and second Birk's law parameters that describe the quenching process for heavily ionizing particles. 
The parameters used for $\alpha$-particles, $\gamma$-rays and electrons, and the light yield were obtained using the LS Screener calibration data~\cite{Haselschwardt:2018}; external $^{137}$Cs and $^{228}$Th $\gamma$ sources were used to fix $Y$ for all particles and to determine $kB$ for $\gamma$/$e^-$, and a $^{220}$Rn source was bubbled through the GdLS to obtain $kB$ and $C$ for $\alpha$-particles. Parameters for protons were taken from~\cite{VonKrosigk:2015yio}, and all are shown in Table~\ref{tab:LSScintParams}. 

\begin{table}
\caption{Parameters used for various particles in production of scintillation photons in the GdLS of the Outer Detector.  \label{tab:LSScintParams}}
\centering
\resizebox{0.48\textwidth}{!}{
\begin{tabular}{c c c c} \hline 
\multirow{2}{*}{\textbf{}} & \textbf{$Y$} & \textbf{$kB$} & \textbf{$C$} \\
& \footnotesize \textbf{(photons/MeV)} & \footnotesize \textbf{(g/MeV/cm$^2$}) & \footnotesize \textbf{(g/MeV/cm$^2$)$^2$} \\\hline  
	$\alpha$  & $9\times10^{3}$ & $4.63\times10^{-3}$ & $1.77\times10^{-6}$\\
    $\gamma$/$e^-$   & $9\times10^{3}$  &0.03  & 0 \\
    Proton  &$9\times10^{3}$   & $8.26\times10^{-3}$ & 0 \\ \hline  
\end{tabular}}
\end{table}

The number of photons calculated using Eq.~\ref{eq:Birks} is smeared using a Gaussian ($>10$ photons) or a Poisson ($<10$ photons) distribution. Photons are produced at positions scattered around the energy depositing track, and with a time distribution according to fast and slow time constants that are chosen based on particle type. Once produced, optical photons may be absorbed by the wavelength shifter within the scintillator, after which they have a wavelength-dependent probability to be re-emitted at a longer wavelength~\cite{Haselschwardt:2018}.

\begin{figure}[h]
\includegraphics[width=0.9\linewidth]{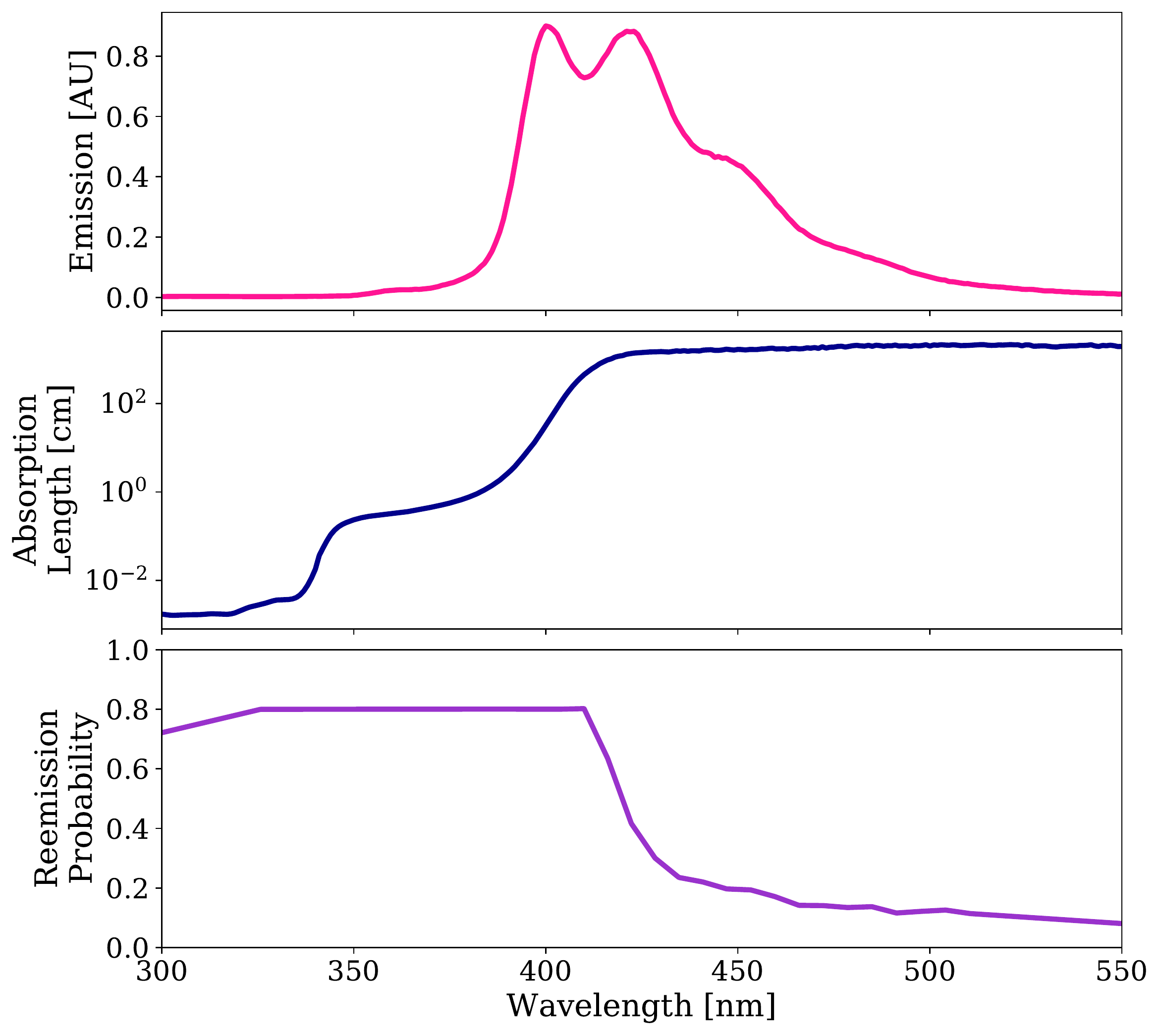}
\caption{Emission spectrum probability of optical photons (top), absorption lengths (middle) and re-emission probability (bottom, plotted as a function of the wavelength of the absorbed photon) of the liquid scintillator used in simulation of the LZ OD~\cite{Haselschwardt:2018}. \label{fig:LSOptProps}}
\end{figure}

The three key optical properties of the LS are the emission spectrum, the absorption spectrum and the re-emission probability; these are shown as implemented for optical simulation of the OD in Figure~\ref{fig:LSOptProps}. The shifted spectrum was sampled independently of the absorbed photon wavelength, with the final intensities tuned to measurements made on the liquid scintillator~\cite{Haselschwardt:2018}.

\subsubsection{Photon Tracking} 
\label{sec:optical}
Photons produced through interactions in the xenon and liquid scintillator are tracked to give the measured signal. A number of parameters affect the light collection efficiency. Most notable are the PTFE reflectivity in liquid and gaseous xenon, for which the latest measurements are used \cite{PTFE}, and the properties of the liquid xenon: both the refractive index \cite{refractive} and Rayleigh scattering length \cite{SEIDEL2002189} are encoded with dependence on the photon wavelength and the thermodynamic properties of the medium. The UNIFIED optical reflection model \cite{UNIFIED} is adopted for all surfaces, and the reflectance and roughness are specified for each. The estimated PTFE coverage of a surface is taken into account when assigning its reflectivity.

Optical tracking in \geantFour was found to be computationally expensive, given the millions of resulting S2 photons routinely generated in background simulations. Simulations with full optics were seen to be many-fold slower than their energy-deposition only counterparts, with the ray-tracing estimated to consume $>$95\% of the CPU time. This hinders the production of large-scale datasets, which are needed for both background estimates and analysis development.

Solutions to this problem involve high statistics approaches. One uses detector-averaged quantities and computes signal sizes via the fastNEST package, described in Section~\ref{sec:NEST}. These quantities include the gains $g1$ and $g2$, which stipulate the number of photons detected per S1 photon and per ionization electron, respectively. Optical simulations, in which VUV photons are emitted from random positions in both the liquid bulk of the TPC and the gas gap between phase boundary and anode, are used to inform the values chosen for $g1$ and $g2$.

These simulations can be further analyzed to extract the hit pattern and times of arrival on the PMT arrays, with output encoded into maps giving the probability of each outcome. This provides information as to how this light is distributed, but there is a trade-off between map resolution and size that makes this a less viable approach for large regions of interest. Fortunately, the S2 signal, which typically dominates over the S1, lends itself to parameterization due to a relatively small volume of origin i.e. the aforementioned gas gap. Suspending tracking of the S2 photons and using the map formulation speeds up the simulations by a factor of 20.

\subsection[Simulating Electronics: Detector Electronics Response]{Simulating Electronics: \\ Detector Electronics Response}
\label{sec:DER}
The Detector Electronics Response (DER) is a software package designed to simulate the PMT signal generation and the subsequent signal processing done by the analogue front-end electronics and digitizers of LZ. It reads in raw photon hits from BACCARAT to create mock digitized waveforms, organized and written in an identical format to output from the planned data acquisition system (DAQ). These can be passed through the LZ Analysis Package (LZap), which performs pulse and event reconstruction, to provide practice data for analysis.

The PMT model considers each photon hit in turn: it first decides whether a photon is detected based on the efficiencies of photoconversion on the photocathode and collection of the emergent photoelectron(s), collectively handled as a quantum efficiency (QE) parameter. This QE is taken to be uniform across the photocathode, with the incidence on it, i.e. the photon transport through the PMT window, handled in BACCARAT.

\begin{figure}[htb]
\begin{center}
\includegraphics[width=8.5cm]{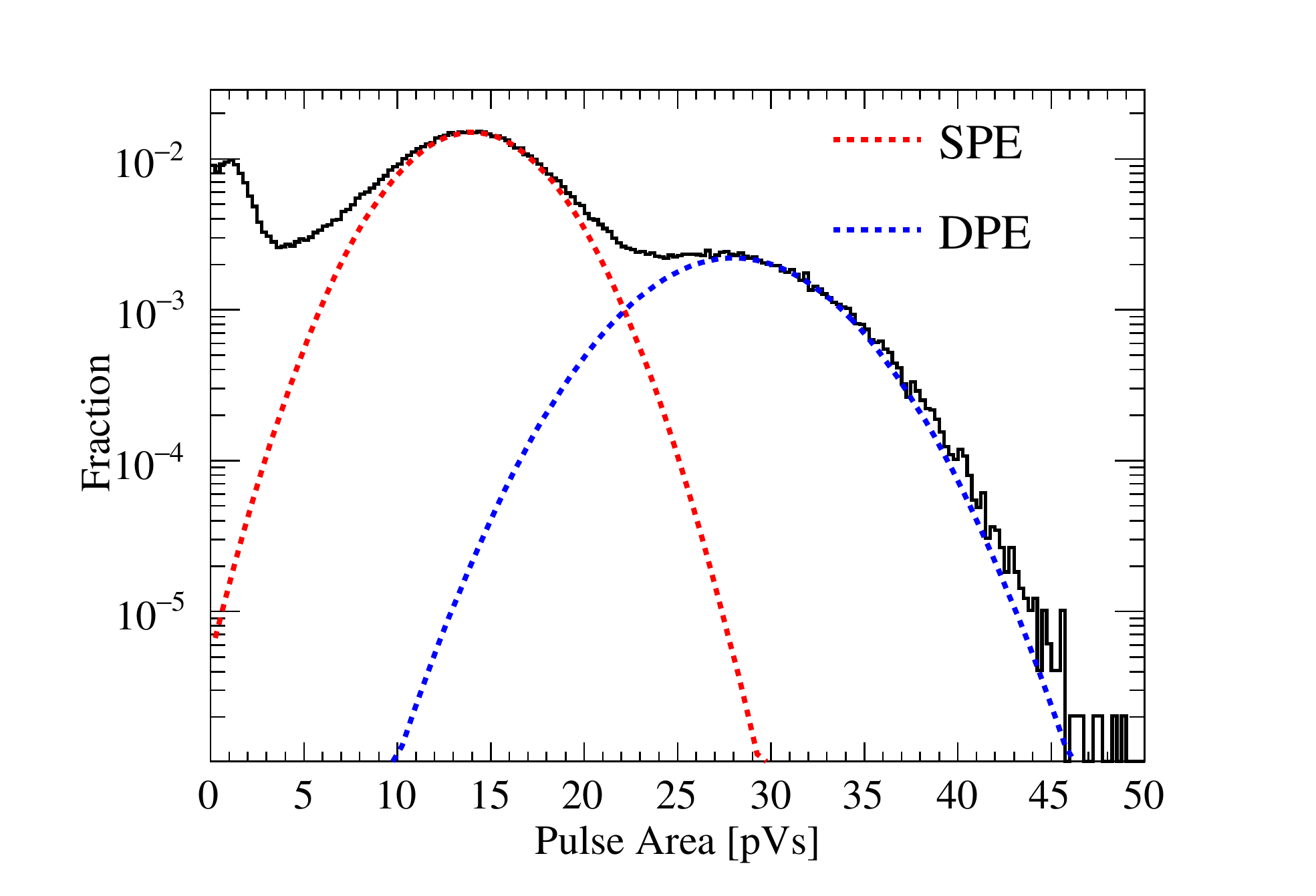}
\caption {Simulated distribution of single photon response pulse areas for a typical R11410-2 Hamamatsu PMT for 175\,nm photons. The single photoelectron (SPE) peak is visible, followed by the DPE peak at roughly 20\% the height \cite{LOPEZPAREDES201856}. First dynode collection leads to the first, low pulse area peak, whilst other effects lead to undersized pulses that govern the peak-to-valley ratios.}
\label{fig:sphReponse}
\end{center}
\end{figure}

The single photon response varies, with the gain and time delay of the signal Gaussianly-distributed about the average multiplication from the dynode chain, and the time of flight of the electrons between each stage of it, respectively. In addition to the standard scenario of a single photoelectron emerging from the photocathode and propagating through the entire dynode chain, several other responses are also accommodated by the model, for which the gain and transit must be separately considered:
\begin{itemize}
    \item The double photoelectric (DPE) effect - if the incident photon is energetic enough, two photoelectrons may be liberated from the photocathode instead of one.
    \item First dynode conversion - the photon may be transmitted through the photocathode and instead photoconvert on the first dynode, thus missing one of the multiplication stages.
    \item Other undersized signals - photoelectrons can inelastically scatter off the first dynode, leading to a delayed undersized response. Additionally, depending on their trajectories and the internal electric fields, photoelectrons may miss the first dynode and collect on the second dynode instead.
\end{itemize} 
The probability of each mode, as well as the QE, may be individual to each PMT, and dependent upon the wavelength of the impingent photon. An example of the single photon response distribution for an LZ TPC PMT is given in Figure~\ref{fig:sphReponse}. The response has been validated against measurements in \cite{LOPEZPAREDES201856}.

Photoelectrons may ionize trace amounts of gas as they accelerate through the PMT, with the ions then stimulating new photoelectron cascades. These afterpulses have a time delay distribution relative to the primary signal and a defined size dependent upon the type of ion. The simulation encodes a variety of common ions, including xenon that may leak into the PMT body. The assumed per ion afterpulsing rates are taken from room temperature measurements of each PMT.

Signals may be produced in PMTs that are independent of photon hits. These so-called dark counts arise from spontaneous thermionic emission from the photocathode or initial stage dynodes, and look identical to a single photoelectron pulse. The dark rate may be specified separately for each PMT in the simulation, and is typically dependent on temperature. The default values used for this effect (as well as QE, DPE and gain) for each PMT in the DER are informed by tests conducted at the expected LZ operating temperature of 175\,K. The average measured dark noise rate for TPC PMTs is 34 Hz, with a spread of approximately 10 Hz either side.

Analytical models are included for the cabling, dual-gain amplifiers and digitizers, based on extensive circuit analysis to derive frequency-dependent transfer functions for each device. These are then approximated by a series of digital filters and gains, which model how the pulses are scaled and shaped in the time-domain \cite{derpaper}.

Lastly, the DER also captures the LZ data extraction and sparsification schemes i.e. the logic to decide when to save data and how to build events. The software presently implements a finite impulse response filter that can be tuned and passed over any waveform, following the prescription used for LUX \cite{LUXTrigger}. The threshold and coincidence requirements for an event trigger based on the output from applying this filter can be configured. The simulations are thus useful in assessing trigger efficiencies and event data volumes for science data and calibration sources.


\section{Simulations of Background Sources}
\label{sec:Backgrounds}
\subsection{Overview}
\label{sec:overview}
All materials used in the detector construction contain some level of naturally occurring radioactive isotopes, such as $^{238}$U, $^{232}$Th, $^{40}$K and $^{60}$Co, all of which contribute to the electron recoil (ER) background component of the experiment via the emission of $\gamma$-rays and electrons that may interact and deposit energy in the active xenon volume (Section~\ref{sec:ERBackgrounds}). Moreover, the first two are also responsible for the production of neutrons through spontaneous fission and ($\alpha,n$) reactions in the materials (Section~\ref{sec:NRBackgrounds}), leading to nuclear recoils (NRs) in the xenon. These NRs mimic the expected WIMP signal if singular in the TPC and not observed with an associated energy deposition in the veto volumes.  

Neutrons and gammas are also produced in the cavern walls surrounding the experiment, originating either from uranium and thorium decay chains in the rock formations (Section~\ref{sec:rockgammas}), or induced by atmospheric muons passing through the rock and detector (Section~\ref{sec:muonneutrons}). The vetoes are critical in suppressing both of these contributions. The neutron flux from radioactivity in rock is efficiently suppressed by at least 6 orders of magnitude by a hydrogeneous shielding (water and scintillator) \cite{LZ-TDR} and will not be considered here.

A fraction of the $^{222}$Rn ($^{220}$Rn) in the decay chain of $^{238}$U ($^{232}$Th) present in materials and residual dust on surfaces in direct contact with liquid or gaseous xenon will escape and mix in the xenon, contributing to the ER background (Section~\ref{sec:Radon}). Plate-out (surface deposition) of $^{210}$Pb (in the $^{222}$Rn chain) on all surfaces due to exposure to air during manufacture and assembly is also considered (Section~\ref{sec:wall}) and leads to an additional NR background due to ($\alpha,n$) reactions or recoiling nuclei in $\alpha$-decays if an alpha is lost in a wall or a dust particulate. Daughters of this $^{210}$Pb, in particular the $\beta$ emitter $^{210}$Bi, may constitute an additional source of ER background if they detach from the internal TPC walls and mix in the xenon. 

To mimic the preliminary selections that will be applied to LZ data for a WIMP search, simulated events not meeting all of the following criteria (hereafter called `standard cuts') are automatically rejected (see Ref. \cite{LZ-sensitivity} for details):
\begin{enumerate}
\item{Single scatter - $\sigma_{r} <$ 3 cm and $\sigma_{z} <$ 0.2 cm, where $\sigma_{r}$ and $\sigma_{z}$ are the energy-weighted standard deviations of hit positions in radial and vertical directions respectively. These cuts are based on the expected position reconstruction resolution from previous experience with LUX data \cite{Akerib_2018}.}
\item{Fiducial events - those that occur within a central cylinder of xenon, defined as having energy-weighted positions with a radius $r <$ 68.8 cm (from the center of the active xenon volume) and a vertical coordinate of 1.5 $< z <$ 132.1 cm (counted upwards from the cathode, and where $z=146.1$~cm is the liquid surface). The fiducial mass of LXe after this cut is about 5.6~tons.}
\item{Non-vetoed events - any interactions in the xenon volume that are not within 500~$\mu$s (800~$\mu$s) of a total energy deposition greater than 200~keV (100~keV) in the GdLS (LXe skin). The choice of the time window for anti-coincidence with the OD is driven by the neutron capture time (a non-negligible fraction of neutrons are captured on hydrogen in acrylic tanks) but limited by a random coincidence rate in the OD to reduce dead time. The time window for the LXe skin is determined by the maximum drift time of the TPC.}
\end{enumerate}
In addition, results are often narrowed to the WIMP energy region of interest (ROI): 1.5–6.5 ~keV$_{ee}$ for ERs and 6–30~keV$_{nr}$ for NRs.

\subsection{Gamma-Rays, Betas and Alphas}
\label{sec:ERBackgrounds}

\subsubsection{Decay Chains and Single Decays}
\label{sec:DecChainGen}
Materials used during the construction of the LZ detector will contain trace amounts of radioactivity, even after a campaign of screening and material selection ~\cite{Akerib:2017Ti}. These radioactive sources constitute a major part of the LZ background~\cite{LZ-sensitivity} and their simulation is therefore crucial in any background estimation. 

Gammas and betas from the decays of radioactive isotopes are modeled using generators within the BACCARAT framework. Alpha particles are also simulated, but they are only relevant for sources inside xenon due to their short range in most materials. The contribution of alphas to the neutron background is modeled separately, and is discussed in more detail in Section~\ref{sec:alphaN}. 

The \geantFour code provides functionality to simulate radioactive decays, using data libraries from the Evaluated Nuclear Structure Data File (ENSDF) \cite{tuli1996evaluated}, which describes the nuclear decays, and from the Livermore Evaluated Atomic Data Library (EADL) \cite{liv2}, which describes any subsequent atomic transitions. Two generators have been written that modify the \geantFour algorithm in several ways to optimize background simulations.

The first lets entire decay chains, or single decays within that chain be simulated. In the former case, the parent isotope and all subsequent decays are processed within a single \geantFour `event'. In the latter scenario, this event is split into `sub-events,’ one for each decay, to provide a clearer association between the decay parent, marked as the progenitor of the sub-event, and the energy depositions by betas/gammas in the detector volumes. This allows for the renormalization of (sub-)event rates in post-processing to reflect measured source activities, which is particularly important in cases where the chain is not in equilibrium. Time is reset for each sub-event such that particle interactions are recorded with sufficient timing precision to ensure data selections can be made based on the relative timing of events in the xenon TPC and veto regions. This is the preferred generator for studies assessing the background contribution of a given decay chain.

The second generator takes as an input the age of a given radioactive source and uses recursive solutions to Bateman equations ~\cite{radsrc} to compute the relative populations of radioisotopes in the corresponding decay chain. The goal is to generate events with realistic starting time stamps, given both the source activity and source age. This accurately mimics the real-time acquisition of data, and enables information from timing or position correlations in the detector to be identified using simulations. This generator has therefore been used extensively in MDCs, where relative timing is paramount and data sets contain events from a mixture of different sources. A similar approach has been used in Ref.~\cite{kareem2011}.

For both generators, special consideration is also given to radioactive decays with short half-lives, for example $^{212}$Po in the $^{232}$Th decay chain and $^{214}$Po in the $^{238}$U decay chain. Given the time duration of an LZ event is about 1~ms, as limited by the maximum drift time, these fast decays may produce signals that overlap with those from a previous decay, and they are therefore recorded in the same BACCARAT event.

\subsubsection{Gammas from Rock}
\label{sec:rockgammas}
\begin{figure}[htb]
\begin{center}
\includegraphics[width=8cm]{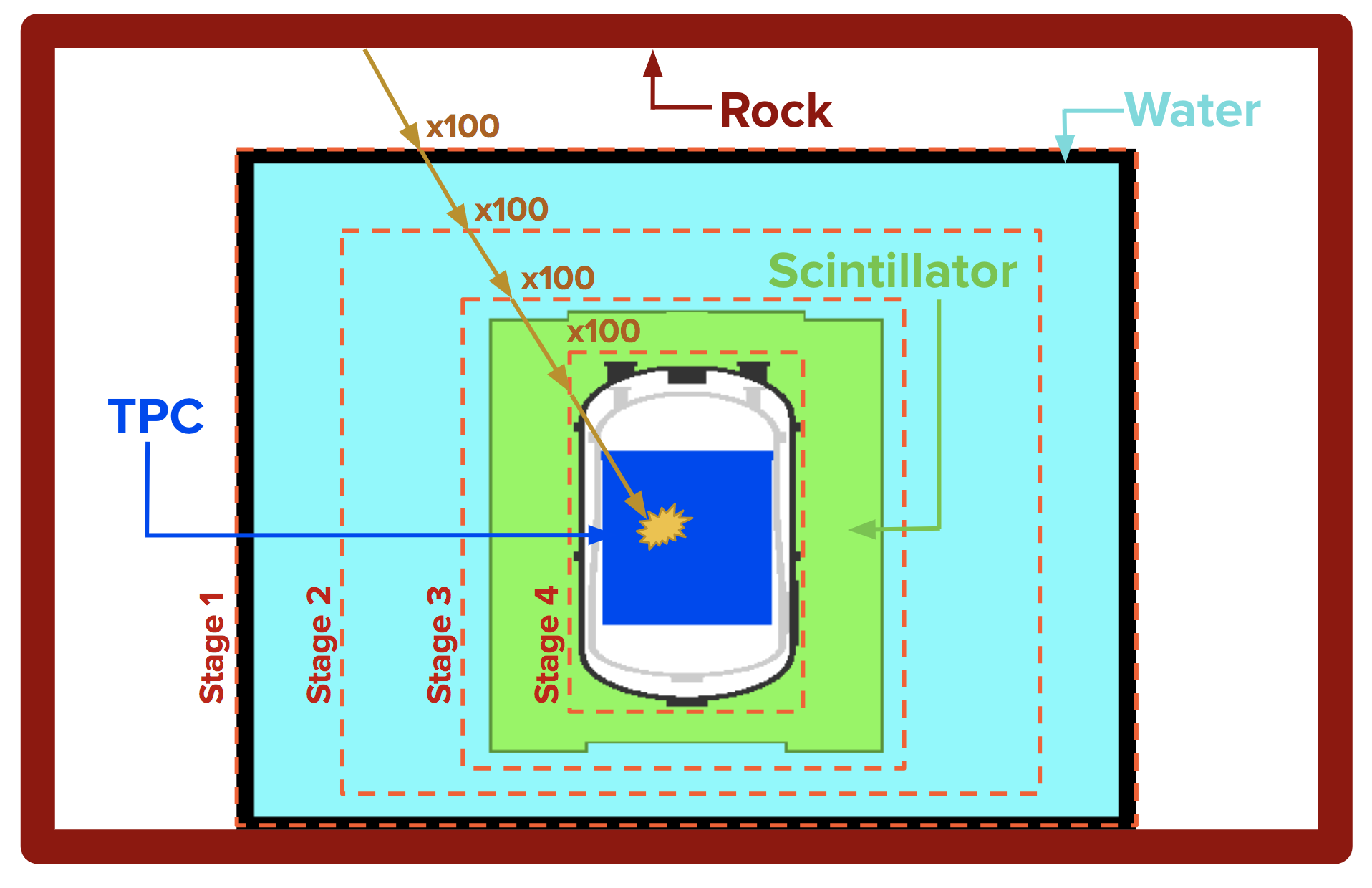}
\caption {A schematic drawing of the event biasing method implemented into BACCARAT allowing the simulations of the cavern rock $\gamma$-rays. The relative dimensions of the cavern, water tank, LS veto, and TPC are not to scale.}
\label{figure:Event-biasing}
\end{center}
\end{figure}

Gamma-rays produced from trace radioisotopes in the rock around the cavern are an external source of background to the LZ experiment. The $\gamma$-flux is effectively attenuated by water, scintillator and the steel pyramid beneath the water tank, with a significant part of the remaining background further suppressed by event selection cuts. Consequently, $\mathcal{O}(10^{15})$ initial decays in the rock need to be simulated to fully characterize this background \cite{David-Woodward-Thesis}. To make this computationally feasible, the simulation is divided into several stages. In the first stage, $\gamma$-rays are produced in the rock using the generator described in Section \ref{sec:DecChainGen}, and those that enter the water tank are saved to file. Concentric cylindrical ‘shells’ are then defined in the simulation geometry, as shown in Figure \ref{figure:Event-biasing}. In successive stages of the simulation, $\gamma$-rays are transported to the next shell boundary, and those that survive are passed to the next simulation stage to be re-propagated multiple times with the same initial position and momentum. Using this scheme with $n$ shells, each with a multiplication factor $m_{i}=100$, the starting number of decays is effectively increased by a factor $f_{b} = \prod_{i=1}^{n} m_{i}$. The systematic uncertainty introduced by this simulation setup dominates the overall uncertainty in the final result, and will be discussed later in this section.  

\begin{figure}[htb]
\begin{center}
\begin{tikzpicture}
    \draw (0, 0) node[inner sep=0] {\includegraphics[width=9cm]{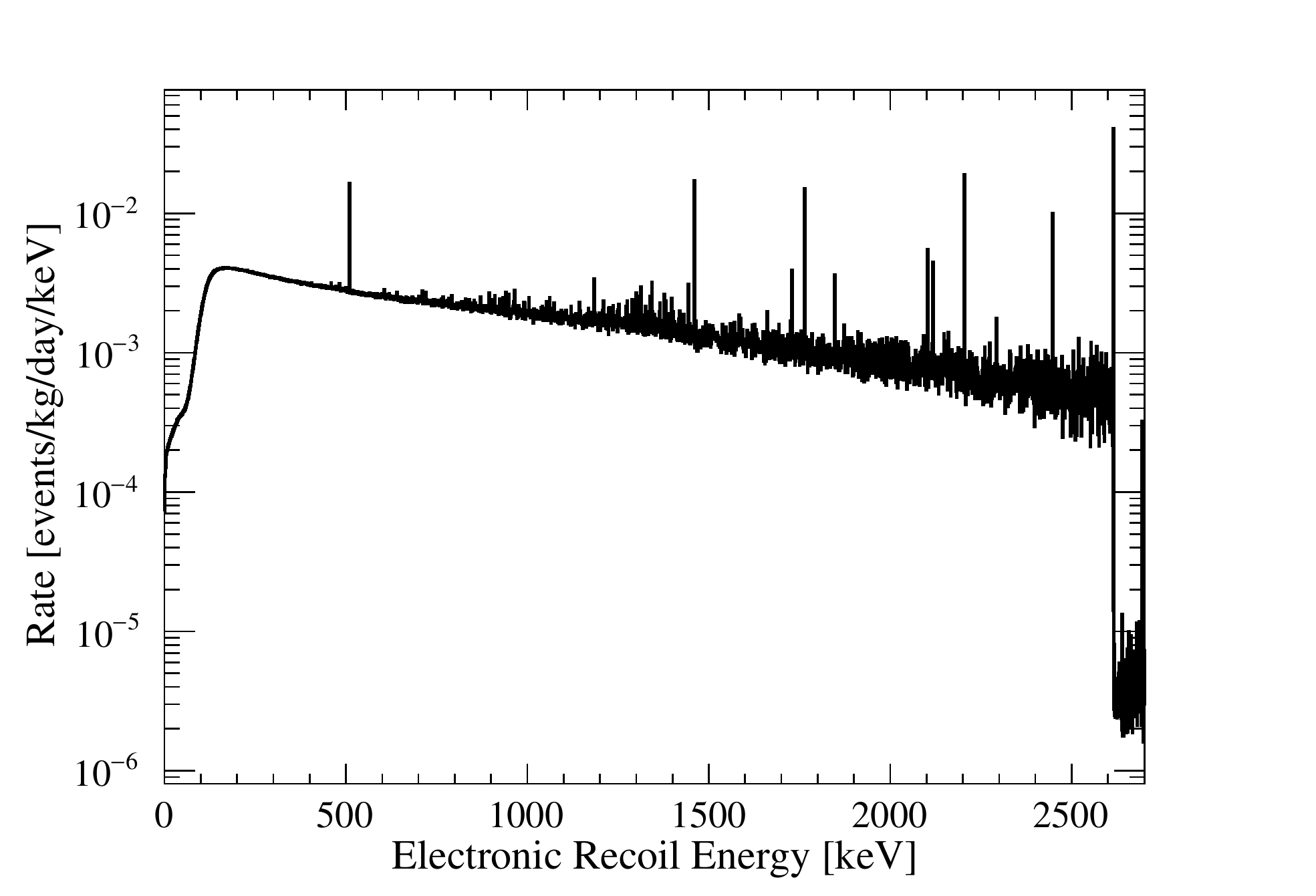}};
	\draw (-1.0, -0.8) node[inner sep=0] {\includegraphics[width=3.78cm]{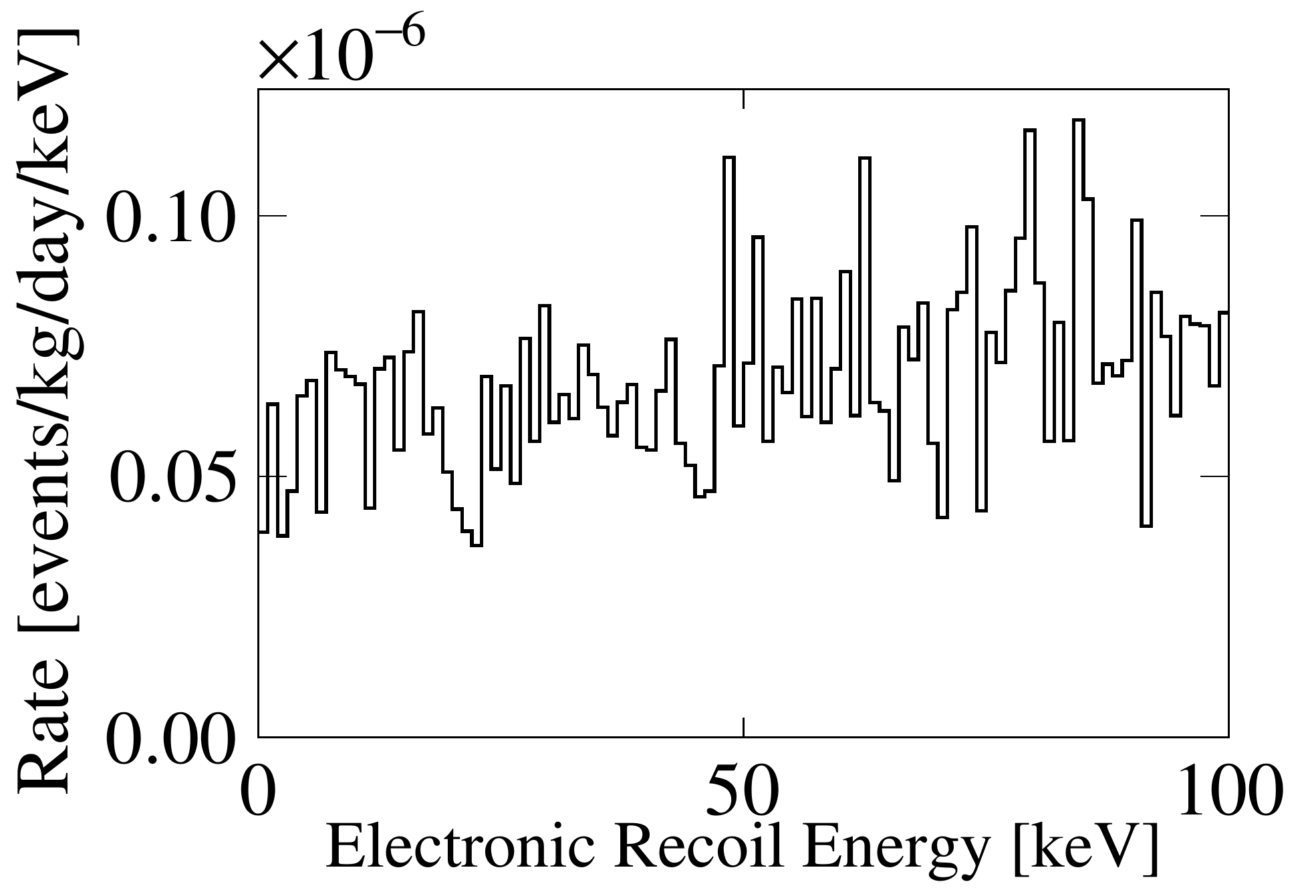}};
\end{tikzpicture}

\caption {The energy spectra of simulated rock gamma events located within the TPC before the application of standard analysis cuts (Section~\ref{sec:overview}). The insert presents the energy spectrum at 
low energies (0-100~keV) and after cuts have been applied.}
\label{figure:Gamma-spectra}
\end{center}
\end{figure}

\begin{figure}[htb]
\begin{center}
\includegraphics[width=8cm]{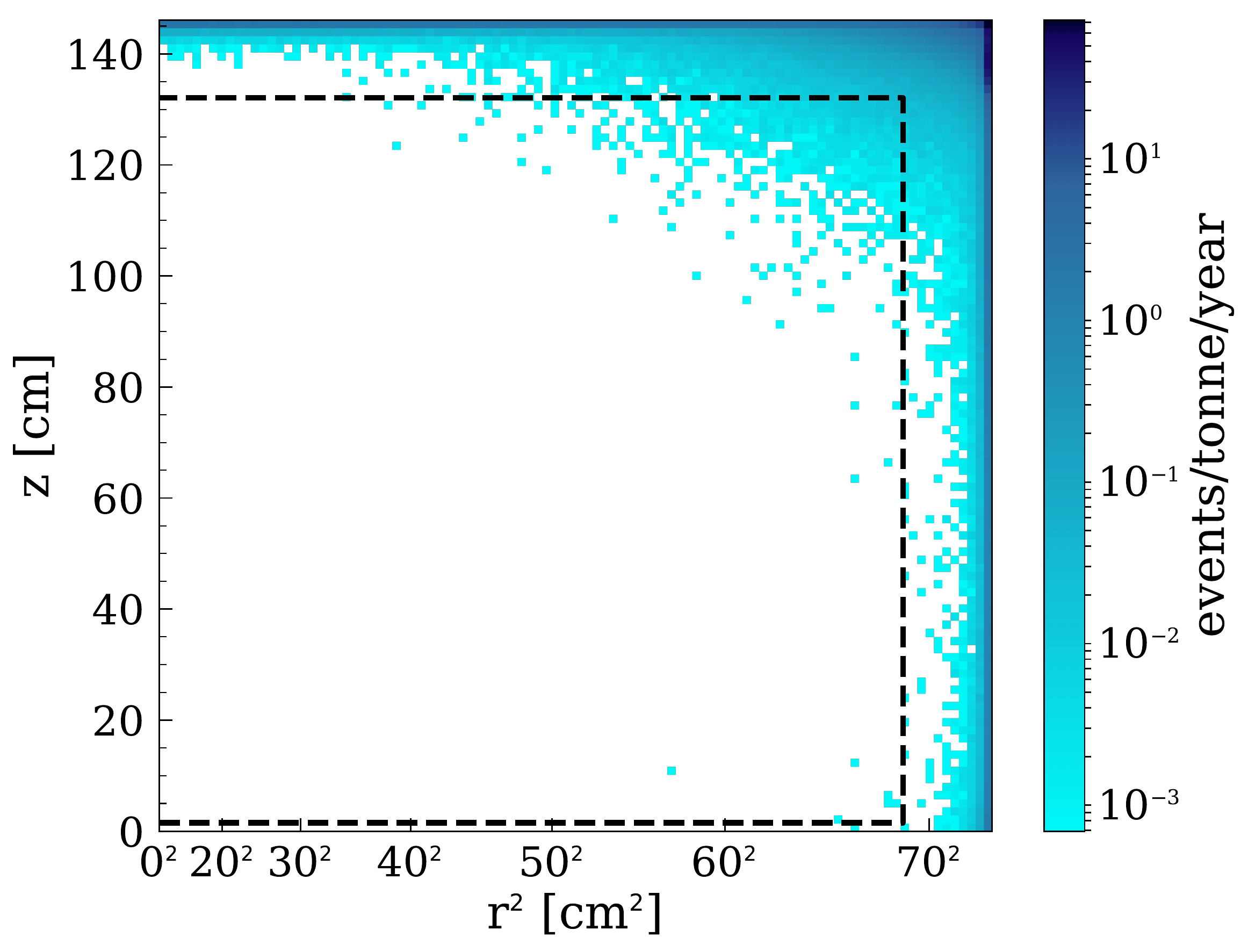}
\caption {Position distribution of single ERs from rock gammas within the LXe for 1.5-6.5~keV energy depositions.
Outer detector and xenon skin vetoes have been applied. The black dashed line illustrates the 5.6 tonne fiducial volume. The cathode is positioned at $z=0$.}
\label{figure:Event-distribution}
\end{center}
\end{figure}

Simulations of $^{232}$Th, $^{238}$U and $^{40}$K decays were undertaken as their daughters dominate the cavern wall $\gamma$-rays. Full decay chains for all isotopes were simulated assuming secular equilibrium and normalized to the following activities taken from the measurements of gamma-ray spectra in the cavern: $^{232}$Th - 13 Bq/kg, $^{238}$U - 29 Bq/kg and $^{40}$K - 220 Bq/kg \cite{Shaw:2019}. Figures \ref{figure:Gamma-spectra} and \ref{figure:Event-distribution} display the resulting energy spectrum and spatial distribution, respectively, of the remaining events within the TPC after standard analysis cuts (Section~\ref{sec:overview}). Figure \ref{figure:Gamma-spectra} also presents the energy spectrum of the remaining events before the standard analysis cuts are applied. Energy depositions in the OD were recorded only after stage 4 (see Figure \ref{figure:Event-biasing}) to reduce the disk space required for the simulation output. This means the result is conservative since some gammas could trigger the OD on their way to the TPC at stage 4.

Figure~\ref{figure:Event-distribution} favors a non-cylindrical shape for the optimum fiducial volume. However, the total background in LZ will likely be dominated by uniformly distributed $^{222}$Rn decay (see Section~\ref{sec:Radon}) whereas the shape and size of the fiducial volume is driven by the wall and detector component backgrounds. There is only a small contribution to the total background from the environment (see Ref.~\cite{LZ-sensitivity} for more details).

The simulations predict 1.81$\pm$0.19 (stat) background events in 1000 live days for a 5.6 ton fiducial mass after standard analysis cuts and before any ER/NR discrimination. Further cuts are also made to select events with S1 signals present in three individual PMTs, and with size greater than 20 detected photons. This approximately corresponds to energy deposits in the range 1.5 - 6.5 keV$_{ee}$, the WIMP ROI. The event rate from cavern walls is sub-dominant to the internal backgrounds, highlighting the effectiveness of the water/scintillator and steel pyramid shielding.

The systematic uncertainty of these simulations is dominated by a potential biasing of the results when propagating surviving gamma-rays multiple times. However, the consistency of the results has been tested by running simulations several times with different positions of the surfaces for the individual stages, with the conclusion that the results are consistent within 20\%.

\subsubsection{Radon}
\label{sec:Radon}
Simulations of radon and its daughters inside the liquid xenon target have been carried out similarly to other decay chains, with the primary particles as ions of $^{222}$Rn or $^{220}$Rn. Radon is emanated from detector materials and, despite its expected low concentration in the xenon thanks to purification \cite{LZ-TDR}, constitutes a significant background to dark matter searches in LZ. This background is dominated by the “naked” beta decay of $^{214}$Pb to $^{214}$Bi in the $^{222}$Rn sub-chain. The $^{214}$Bi beta decay will be tagged by the subsequent $^{214}$Po alpha decay, whilst other transitions to excited states can be identified by coincidences between betas and gammas. Similar considerations apply to the $^{220}$Rn sub-chain. ER interactions in the TPC from Bi-Po delayed coincidences were assumed to be captured within a single event since only 3.4\% of $^{214}$Po decays will happen outside of the time window of 800 microseconds (corresponding to the expected maximum drift time) opened by the preceding $^{214}$Bi decay. 

$^{210}$Pb and its progeny's decays are not included in the simulations as they are presumed to be removed by the continuous recirculation and purification of the xenon. For those daughters that adhere to the surfaces, and therefore cannot be eliminated via purification, a fiducial volume cut is highly efficient in reducing the background from those decays.

\begin{figure}[htbp]
\begin{center}
\includegraphics[width=9cm]{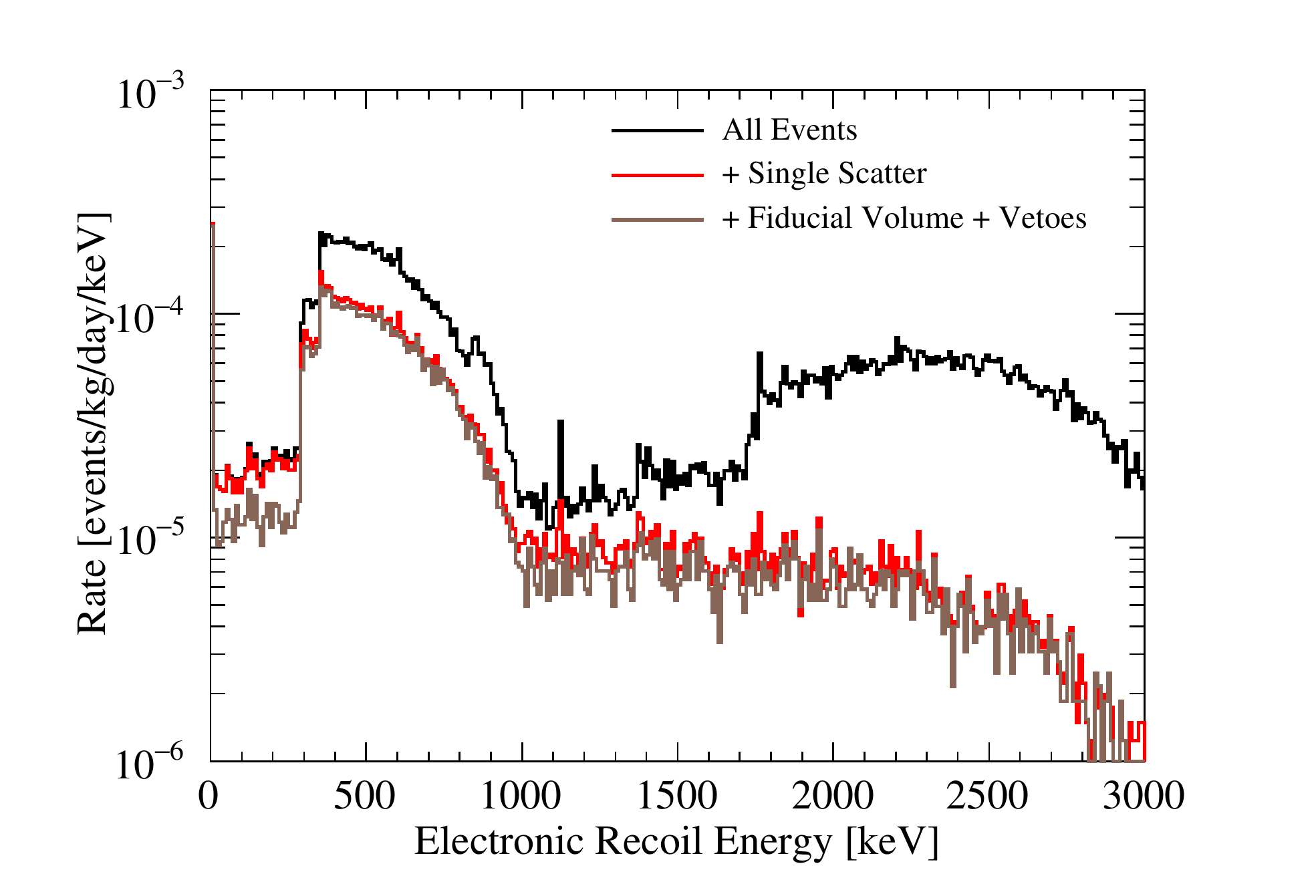}
\caption{Energy spectra of ER events from radon decay. Black -- all events in LXe; red -- events with a single scatter in the TPC; brown -- events after all cuts (including veto and fiducial volume cuts, but without NR/ER discrimination). An initial radioactivity of 1~$\mu$Bq/kg of radon has been assumed for normalization.}
\label{fig:radon}
\end{center}
\end{figure}

Figure~\ref{fig:radon} shows the energy spectra of events from $^{222}$Rn sub-chain before and after standard cuts (Section~\ref{sec:overview}). The radioactivity level of 1~$\mu$Bq/kg of $^{222}$Rn was used to normalise the curves in Figure~\ref{fig:radon}. The LZ background model assumes 1.8~$\mu$Bq/kg of $^{222}$Rn \cite{LZ-sensitivity} based on measured emanation rates, giving 681 events in the fiducial volume in 1000 days. This background is expected to dominate in the science runs of LZ. In the calculation of the background rate in the range of interest (1.5 -- 6.5~keV) the branching ratio of 9.2\% for the decay of $^{214}$Pb into the ground state of $^{214}$Bi was used \cite{TabRad_v8}, instead of the default \geantFour value of 6.3\% (this additional scaling of BACCARAT simulations has not been applied to the curves in Figure~\ref{fig:radon}). Similarly, a 13.3\% probability of $^{212}$Pb decaying into the ground state of $^{212}$Bi was assumed \cite{TabRad_v8}. Events above 1~MeV are very unlikely to be single scatters causing a significant reduction of the event rate at these energies after selecting only single scatters.

\subsection{Neutrons from Radioactivity} 
\label{sec:NRBackgrounds}

\subsubsection{($\alpha,n$) Reactions}
\label{sec:alphaN}
Neutrons emitted from radioactive processes in materials near the LXe target can produce isolated nuclear recoils that can mimic those expected from WIMPs. To simulate neutron backgrounds from radioactivity (the $^{238}$U, $^{235}$U and $^{232}$Th decay chains), BACCARAT uses input neutron spectra calculated with the SOURCES4A package \cite{sources4}.

The SOURCES4A code calculates neutron yields and spectra from spontaneous fission, ($\alpha,n$) reactions and delayed neutron emission due to the decay of radionuclides. Its library contains all alpha emission lines from known radioactive isotopes. The code takes into account the energy losses of alphas, cross-sections of ($\alpha,n$) reactions and the probabilities of nuclear transition to different excited states (excitation functions). We use an option for a thick target, allowing for the calculation of neutron yields and spectra under the assumption that the size of a material sample significantly exceeds the range of alphas. The original SOURCES4A code has been modified \cite{carson2004,lemrani2006,tomasello2008} to extend the energy range of alpha particles to 10 MeV and to include ($\alpha,n$) cross-sections and excitation functions for most isotopes relevant to underground rare event experiments, based either on measurements or on EMPIRE2.19 code \cite{empire}. 

\begin{figure}[htbp]
\begin{center}
\includegraphics[width=8cm]{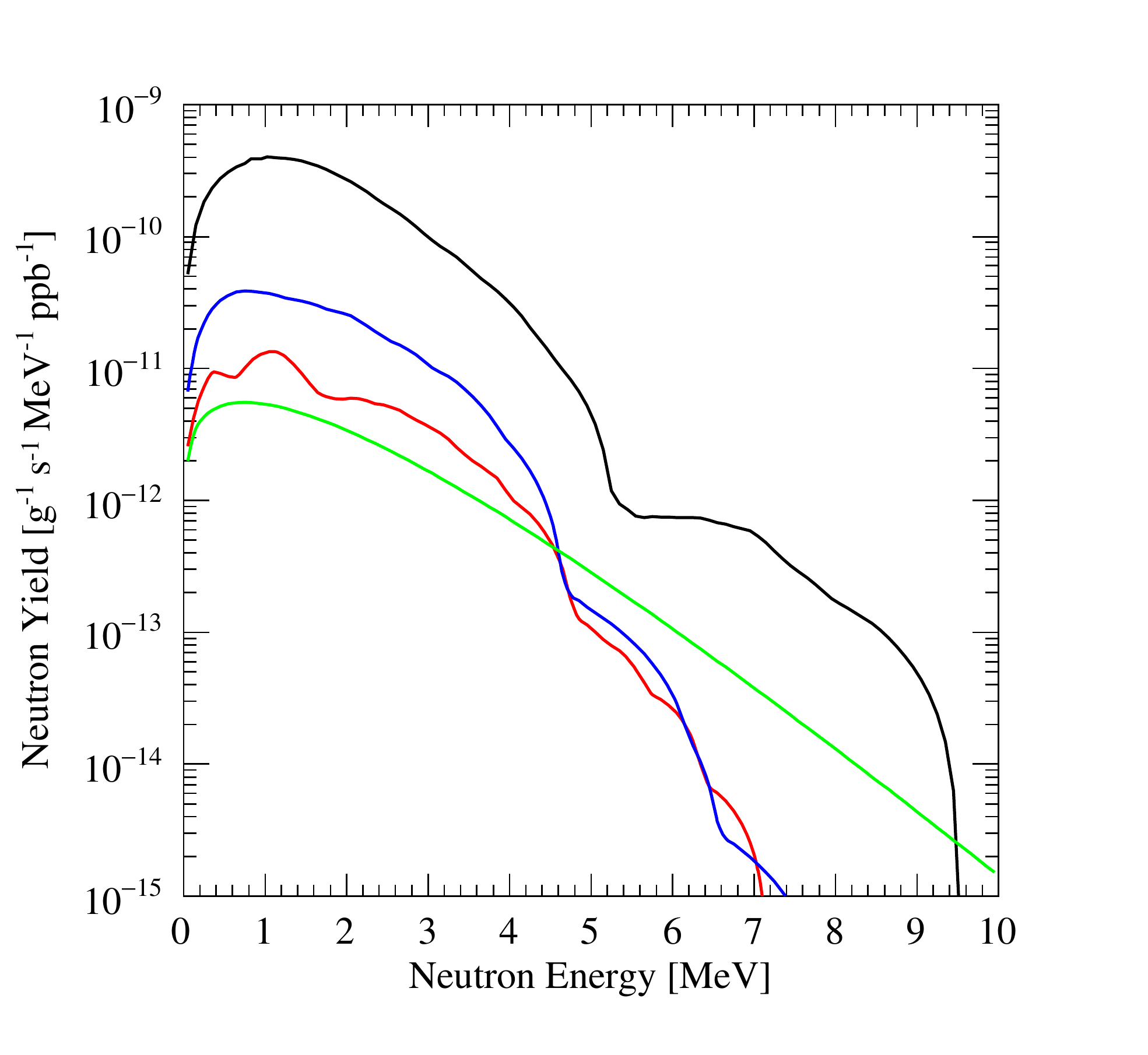}
\caption{Neutron spectra from ($\alpha,n$) reaction from uranium decay chains in equilibrium ($^{238}$U and $^{235}$U are combined together) in 3 materials: black - PTFE (C$_2$F$_4$), blue - ceramics (Al$_2$O$_3$), red - titanium. The green curve shows the spectrum from spontaneous fission (same for all materials).}
\label{figure:sources4}
\end{center}
\end{figure}

The neutron spectra from SOURCES4A are implemented as generators in BACCARAT, allowing any detector component to become a source of neutrons. The measurements of the radioisotope concentrations or decay rates are used to scale the simulation results to predict the background rates. The $^{238}$U decay chain is split into the early (before $^{226}$Ra) and late (starting from $^{226}$Ra) sub-chains, and the $^{210}$Pb sub-chain is calculated separately if required. The $^{235}$U decay chain is not split (due to short lifetimes of all isotopes below $^{235}$U) and is added to the early $^{238}$U decay chain. Figure~\ref{figure:sources4} shows example neutron spectra from PTFE, titanium and ceramics (Al$_2$O$_3$) from the whole uranium chain assumed to be in equilibrium. These materials have been chosen as examples because they either have a high mass or a significant neutron yield per unit activity. Both spontaneous fission and ($\alpha,n$) reactions are shown on this plot but spontaneous fission is not included in the background estimate due to the predicted high efficiency of simultaneous detection of neutrons and gammas from this process (see Section~\ref{sec:SF}).

Appendix~\ref{app:neutron-yields} includes a table with neutron yields as calculated using SOURCES4A and used in the evaluation of backgrounds for LZ.

\subsubsection{($\alpha$,n) Neutrons with Coincident Gammas}
A significant contribution to the neutron-induced NR background in LZ may come from ($\alpha,n$) reactions in the PTFE reflector lining the TPC due to the relatively large ($\alpha,n$) cross section of fluorine \cite{LZ-sensitivity}. Uranium and thorium contaminants in the PTFE will produce alpha decays that cause ($\alpha,n$) reactions. Bulk contamination of PTFE powder with $^{210}$Pb is also expected, thus generating $^{210}$Po that will induce further neutron emission. Lastly, radon daughters are known to plate out on the PTFE surfaces and induce these reactions. 

The $^{19}$F($\alpha,n$)$^{22}$Na reaction often leads to excited states of $^{22}$Na and subsequent de-excitation via $\gamma$-ray emission. Neutrons are usually considered to be produced within `neutron groups’ (nuclear levels associated with gamma emission), $n0$, $n1$, $n2$ etc., corresponding to the state of  $^{22}$Na, with $n0$ as the ground state. The peak of the neutron energy distribution shifts to lower energies as the excitation level of $^{22}$Na increases, and the total emitted $\gamma$-ray energy increases. SOURCES4A provides neutron energy spectra for each $\alpha$-decay and each excited state, but no information on the resulting $\gamma$-cascade. Relative populations of each excited state were calculated using SOURCES4A output for the early chain of uranium, the late chain of uranium, the thorium chain and $^{210}$Po. These were combined with the energy spectra summed over $\alpha$-decays for each excited state and data from Nuclear Data Sheets for $^{22}$Na to produce a PTFE ($\alpha,n$) generator. This generator first chooses a final state for $^{22}$Na using a random number generator and known probabilities of transitions to the ground or different excited states. It then produces a neutron sampled from the corresponding neutron group energy spectrum, and if the chosen state is greater than $n0$, gammas of the appropriate energies and branching ratios will also be generated. Neutrons and gammas are generated isotropically and energy-angular correlations are not taken into account. Angular correlations between neutron and gamma emissions may have only secondary order effects on the efficiency of neutron rejection by simultaneous gamma detection in any active detector volume, due to an almost $4\pi$ coverage of the xenon TPC by the skin and the OD.

\begin{figure}[htb]
\begin{center}
\includegraphics[width=8.8 cm, trim = 40 15 68 10]{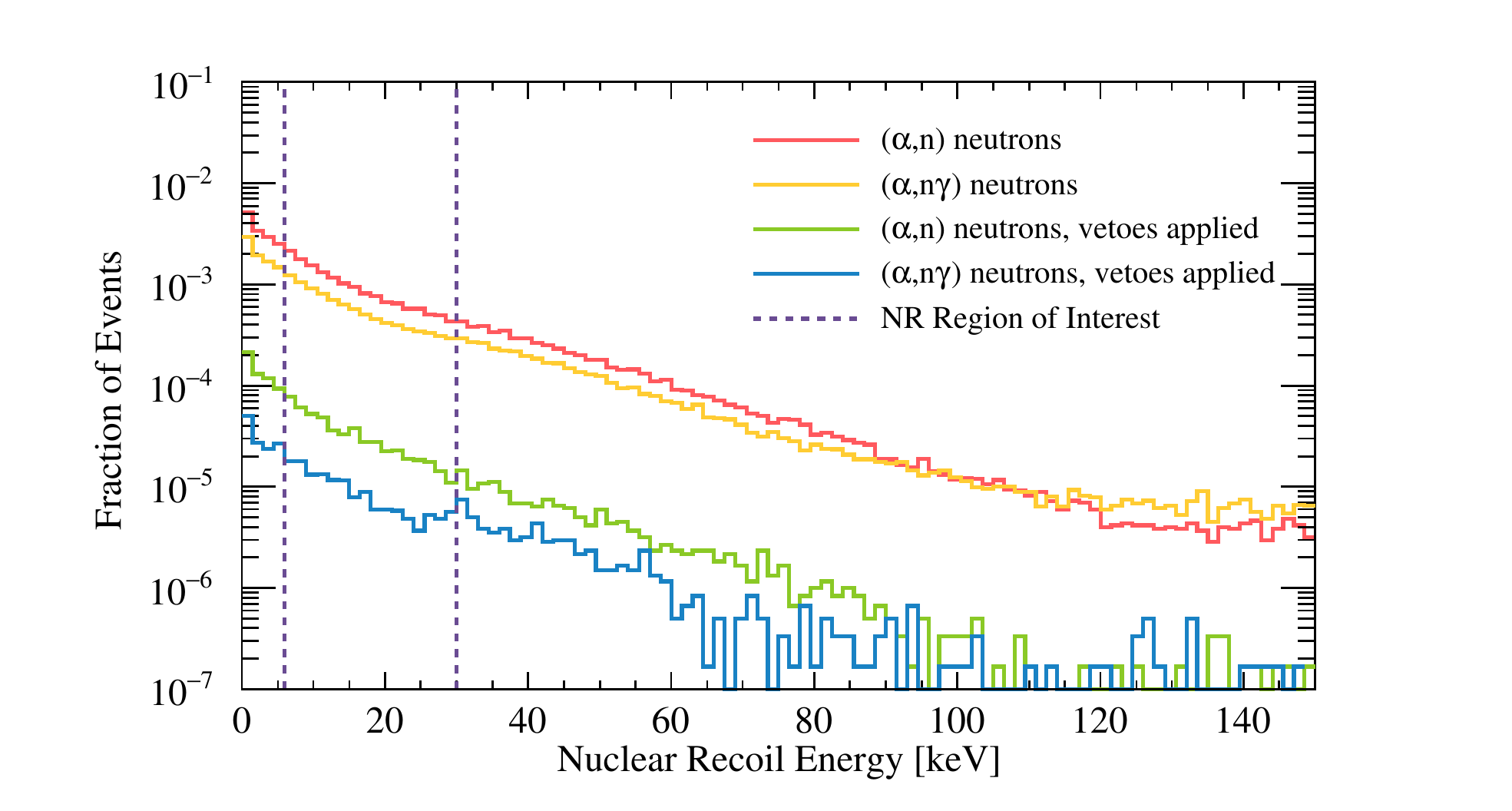}
\caption {Energy spectra of nuclear recoils in the 0--150~keV region produced by single scatter neutrons from ($\alpha,n$) reactions in PTFE, before and after application of vetoes. The red and green histograms use only neutrons from the two generators described in the text, whilst the blue and yellow include coincident $\gamma$-rays and correctly populated neutron groups. The spectra of energy depositions from single scatter neutrons in red and blue differ due to the method of combining ER and NR energies in the liquid xenon in the post-processing of simulations. The blue histogram has more events with a higher energy due to the inclusion of the $\gamma$-rays, which sometimes scatter in the TPC but may not deposit enough energy to cause the event to fail the single scatter cut. The difference between the yellow and green curves demonstrates the increased effectiveness of the vetoes when coincident $\gamma$-rays are included.}
\label{fig:alpha-n-gammas}
\end{center}
\end{figure}

Compared to results from simulations with only neutrons, significantly more events are rejected because of detection of a coincident $\gamma$-ray by the LXe skin
(see Figure \ref{fig:alpha-n-gammas}). Notably, the rejection factor is also better for neutrons emitted from the $n0$ ground state (i.e. without an accompanying $\gamma$-ray). The energy spectrum for the $n0$ group is harder than the one used when summing over all neutron groups. This results in fewer $n0$ neutrons with energies between 0.4–2~MeV, which is the general range required for neutrons to elastically scatter from a xenon nucleus and induce an NR in the WIMP ROI (6–30~keV$_{nr}$). In all, 60\% fewer NRs are now in the WIMP ROI when using 
neutron spectra for individual neutron groups compared with the cumulative spectrum over all final states. As a result of all these effects, the overall cut efficiency increases by 60–80\% with better rejection achieved for chains with the highest energy $\alpha$-decays and therefore the higher probability of $\gamma$-emission (see Table \ref{tab:alphan}).

\begin{table}
\caption{Survival probabilities for all WIMP search background cuts for neutrons produced by ($\alpha,n$) reactions in PTFE. Shown for comparison are single neutrons generated with an energy spectrum summed over all final state neutron groups, and the ($\alpha,n\gamma$) generator, which produces coincident $\gamma$-rays and samples neutron group energy spectra separately. The second and third columns show the fraction of surviving neutrons in the two models, whereas the 4th column shows the ratio of surviving neutrons in the ($\alpha,n\gamma$) generator relative to the original neutron generator. \label{tab:alphan}}
\begin{ruledtabular}
\begin{tabular}{lccc} 
\multirow{2}{*}{\textbf{$\alpha$ source}} & \multicolumn{2}{c}{\textbf{Rejection factor}} & \textbf{Relative} \\
& neutrons & ($\alpha$,n$\gamma$) & \textbf{Suppression} \\ \hline 
$^{238}$U early chain & $1.20\times10^{-4}$ & $4.65\times10^{-5}$ & 0.38 \\
$^{238}$U late chain & $1.08\times10^{-4}$  & $2.00\times10^{-5}$ &  0.19\\
$^{232}$Th chain & $1.04\times10^{-4}$ & $1.95\times10^{-5}$  & 0.19 \\
$^{210}$Po &$1.00\times10^{-4}$ & $3.10\times10^{-5}$ &  0.31 \\
\end{tabular}
\end{ruledtabular}
\end{table}

In the LZ background model \cite{LZ-sensitivity} the plate-out of radon progeny on PTFE surfaces was assumed to be 0.5~mBq/m$^{2}$ (as a target based on rigorous cleanliness program) resulting in 0.05 NRs in 1000 days from misreconstructed near-surface events leaking into the fiducial volume and neutrons from ($\alpha,n$) reactions in PTFE (from $^{210}$Po decay). In addition, 10~mBq/kg of $^{210}$Pb accumulated in the bulk of PTFE during the manufacturing process will give about 0.12 NRs events in the fiducial volume in 1000 days.

Simulations using this PTFE ($\alpha,n$) generator were not used for LZ sensitivity estimates presented in Ref. \cite{LZ-sensitivity}. Since we expect ($\alpha,n$) related background to be lower due to the detection of coincident gammas, the NR background rates shown in Ref. \cite{LZ-sensitivity} are conservative. 

However, the reduction in the total NR background rate due to the detection of coincident gammas from PTFE only, is quite small because of low radioactive contamination of this material.
Nevertheless, this generator is particularly useful for testing and validating procedures of reconstructing coincident gammas and neutrons.

\subsubsection{Spontaneous Fission}
\label{sec:SF}
$^{238}$U spontaneous fission (t$_{1/2}$ = $4.468\times10^9$~years) can contribute to neutron yields significantly in some materials where ($\alpha$,n) yields are low and early U-chain radioactivity is high. A $^{238}$U spontaneous fission event will have near-simultaneous emission of up to 6 neutrons and 20 $\gamma$-rays, with an average yield of about 2.01 neutrons \cite{neutron-mult} and 6.36 gammas \cite{gamma-mult}. In order to evaluate the fission neutron vetoing efficiency of LZ, an event generator was developed that produces multiple neutron and $\gamma$-rays sampled from multiplicity and energy distributions from the Fission Reaction Event Yield Algorithm (FREYA) software~\cite{FREYA:2014}.  Since spontaneous fission is material independent, this generator may be used in any detector component in the simulation.

Simulations of spontaneous fission events in both the cryostat and PMTs were performed and compared to results using neutrons emitted individually with the same energy spectrum. A notable increase in the vetoing efficiency of the outer detector was observed; this effect was expected considering not only the increase in the number of neutrons but also the accompanying $\gamma$-rays of an energetic fission event. The overall rejection power of the outer detector was observed to increase by a factor of 34 for events from the cryostat and 55 for events from the PMTs (see Appendix~A for neutron yields due to spontaneous fission and ($\alpha, n$) reactions from different materials). Spontaneous fission neutrons were therefore considered vetoed with such a high efficiency that they were subsequently removed from the background model. 

\subsubsection[Neutrons in the OD: DICEBOX Neutron Capture Model]{Neutrons in the OD: DICEBOX Neutron Capture Model}
\label{sec:DB}
In pure liquid scintillator, neutrons are captured mainly on protons with subsequent emission of a single 2.22~MeV $\gamma$-ray associated
with the formation of $^{2}$H. When a neutron is captured on $^{155}$Gd or $^{157}$Gd,
the de-excitation process is more complex; there are many excited states of the final Gd isotopes. Energy is released in a de-excitation cascade of $\gamma$-rays, internal conversion electrons, and X-rays.
Modeling of these cascades is complicated and the standard \geantFour toolkit does not provide full and correct description.

We have incorporated the DICEBOX\cite{Becvar:1998} simulation of the de-excitation cascade after neutron capture on $^{155}$Gd  and $^{157}$Gd  to incorporate energy conservation and connect the continuum of states to the well-defined discrete energy levels of excited $^{156}$Gd and $^{158}$Gd.
The DICEBOX package is a nuclear physics software suite that uses a statistical approach to model gamma cascade de-excitation processes. The authors tuned and validated their software by comparing with data taken by the DANCE detector \cite{Baramsai:2013xga, Chyzh:2011zz}, which measured gammas emitted in $^{155}$Gd and $^{157}$Gd neutron captures.

The integration of DICEBOX into BACCARAT was performed via look-up tables for Gd isotopes, with entries showing the amount of energy released in each step of the de-excitation cascade and whether a gamma or internal conversion electron is ejected. When a neutron is captured by either of the two Gd isotopes, an entry from the DICEBOX database is read and post-capture particles are generated according to the DICEBOX calculated probabilities. The binding energy of the internal conversion electrons is subtracted from the electron emission energy, and then an X-ray with the binding energy is added to the post-capture particle list. The directions of the particles in the de-excitation cascade are assigned new random directions without any correlation between them, and a momentum is assigned to the daughter nucleus to preserve momentum conservation.  

\begin{figure}
\begin{center}
\includegraphics[width=0.52\textwidth]{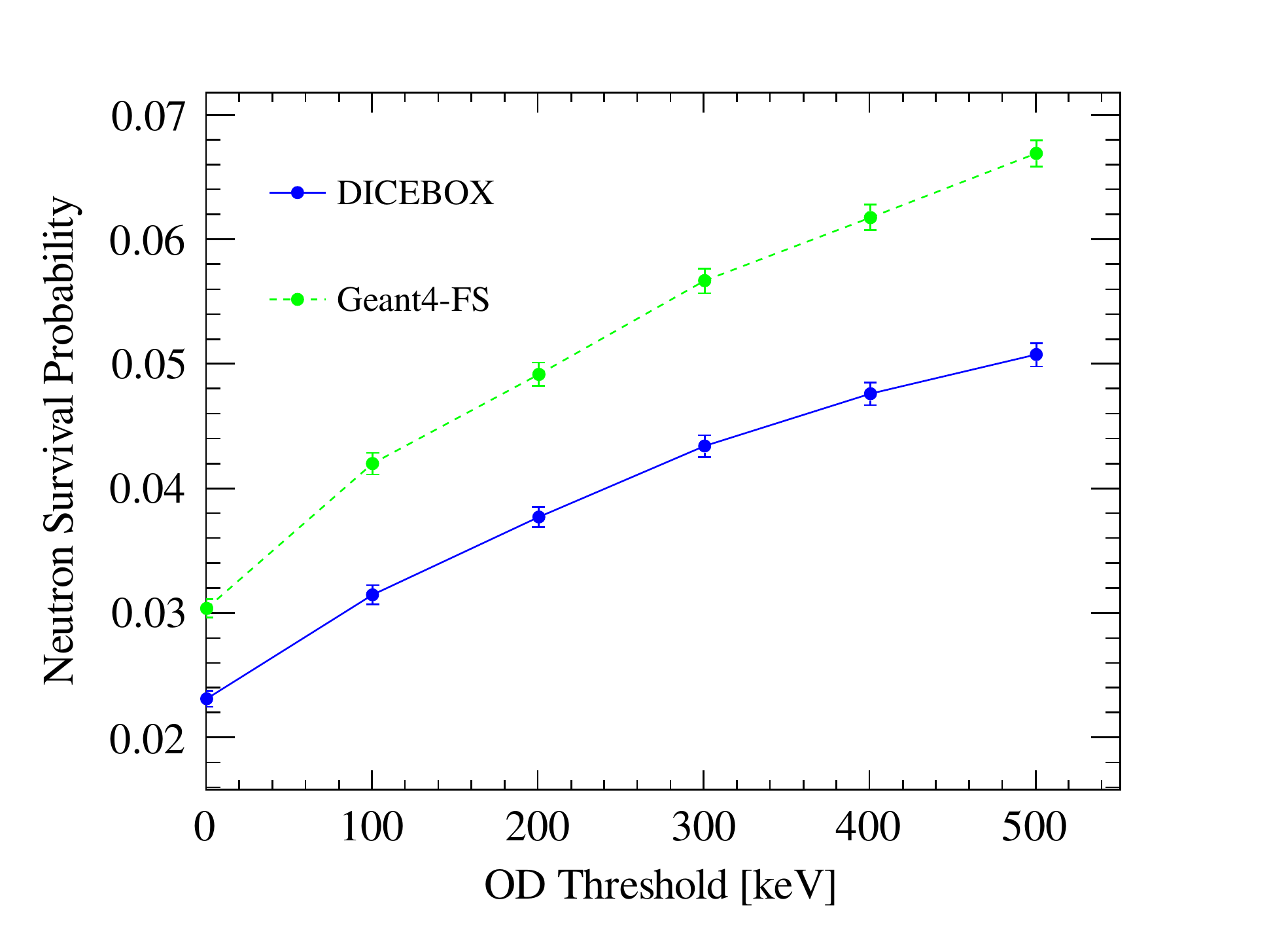}
\caption{The LZ neutron survival probability versus OD energy threshold. The solid blue line shows the result using the DICEBOX simulation for Gd neutron capture, and the dashed green line shows the result from the default \geantFour Final State (FS) neutron capture model. 
The baseline threshold for the OD is 200~keV to avoid false vetoes from the decays of
$^{14}$C, $^{152}$Gd, and $^{147}$Sm in the GdLS. Simulations also assume the energy threshold for the LXe skin of 100~keV.}
\label{fig:inefficiency}
\end{center}
\end{figure}

To evaluate the impact of DICEBOX on the neutron veto efficiency of the OD, about 200 million neutron events coming from LZ detector components were simulated. The yields and the energy  spectra of the neutrons from these components were determined to the best of our knowledge (see Section~\ref{sec:NRBackgrounds}) and the yields were weighted properly when calculating efficiencies. All simulated events include single neutrons only and no coincidences between several neutrons, or neutrons and gammas were considered in these simulations. The neutron survival probability was calculated as the ratio of events surviving all cuts as described in Section~\ref{sec:overview}, to those without skin and OD cuts. Figure~\ref{fig:inefficiency} shows that, due to an accurate treatment of gamma cascades after neutron capture, with an OD threshold of 200~keV, the neutron survival probability evaluated with DICEBOX is 3.8\%, compared to 4.9\% evaluated with the default \geantFour neutron capture model.

\subsection{Muons and Muon-Induced Neutrons}
\subsubsection{Muon model}
Energetic neutrons are produced by atmospheric muons that penetrate through the rock surrounding SURF. Evaluations of the impact of these neutrons must begin with understanding their creation, and therefore the development of a muon flux model.  Atmospheric muons with different energies were propagated through rock with known composition and density using the MUSIC code \cite{music1997,musun2009}. The energy distributions of these muons were recorded at several depths that cover the whole range of distances that muons can cross at different zenith and azimuthal angles before reaching SURF. These energy distributions have then been convoluted with the energy spectra and angular distributions of muons at the surface, with the surface profile taken into account (see Ref. \cite{musun2009} for a detailed description of the procedure). Muon energy spectra and angular distributions obtained this way were recorded and used to sample muons at SURF around the cavern (the MUSUN code \cite{musun2009}). Muons sampled with MUSUN are passed to the LZ software package for further simulations of muon-induced effects. To validate the muon model, the vertical and total muon fluxes were compared with existing measurements in the Davis cavern where LZ will be located. Vertical muon intensity has been measured in early 1980s by the veto system of the chlorine solar neutrino experiment giving the value of $(5.38\pm0.07)\times10^{-9}$ cm$^{-2}$~s$^{-1}$~sr$^{-1}$ \cite{cherry1983}. This can be compared with the LZ model, which gives the vertical muon intensity of $5.18\times10^{-9}$ cm$^{-2}$~s$^{-1}$~sr$^{-1}$ (note that the measured vertical intensity of single muons has been corrected by us to include the reported fraction of multiple muon events). Recently, the total muon flux has been measured in another hall at SURF with a veto system of the Majorana demonstrator. The measured value of $(5.31\pm0.17)\times10^{-9}$ cm$^{-2}$~s$^{-1}$ \cite{mjd-muons2017} is slightly lower than our model prediction for the total muon flux of $6.16\times10^{-9}$ cm$^{-2}$~s$^{-1}$. In both measurements only statistical uncertainties are quoted. Given a (-7+16)\% difference between our model predictions and the two measurements (one for the vertical muon intensity and the second one for the total muon flux), we estimate the accuracy of the model in calculating the muon flux as about 20\%, due primarily to the uncertainty in the rock density. Validation of the muon propagation code was reported in \cite{music1997,musun2009}. The mean muon energy at SURF is calculated to be 283 GeV.

\begin{figure}[htb]
\begin{center}
\includegraphics[width=8cm]{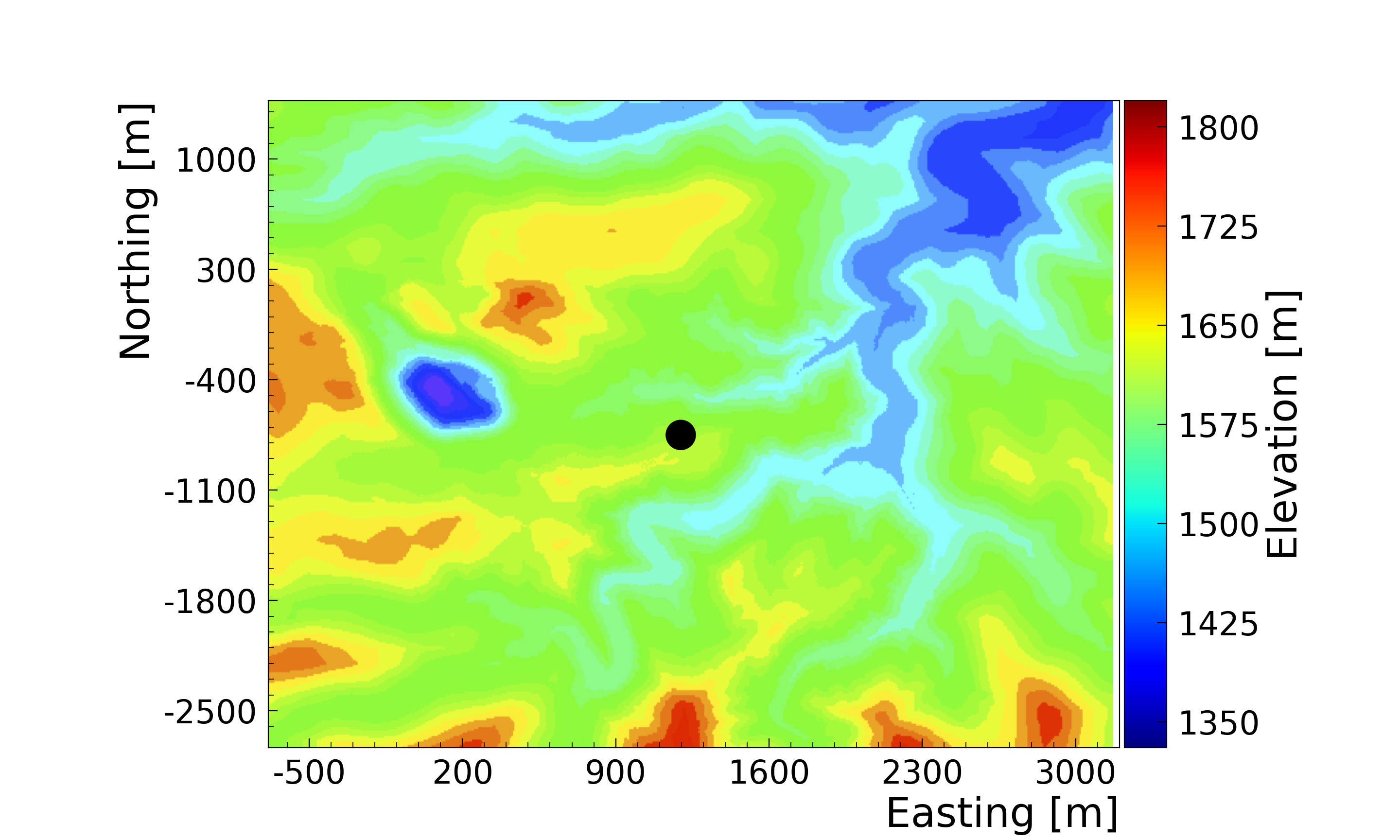}
\includegraphics[width=8cm]{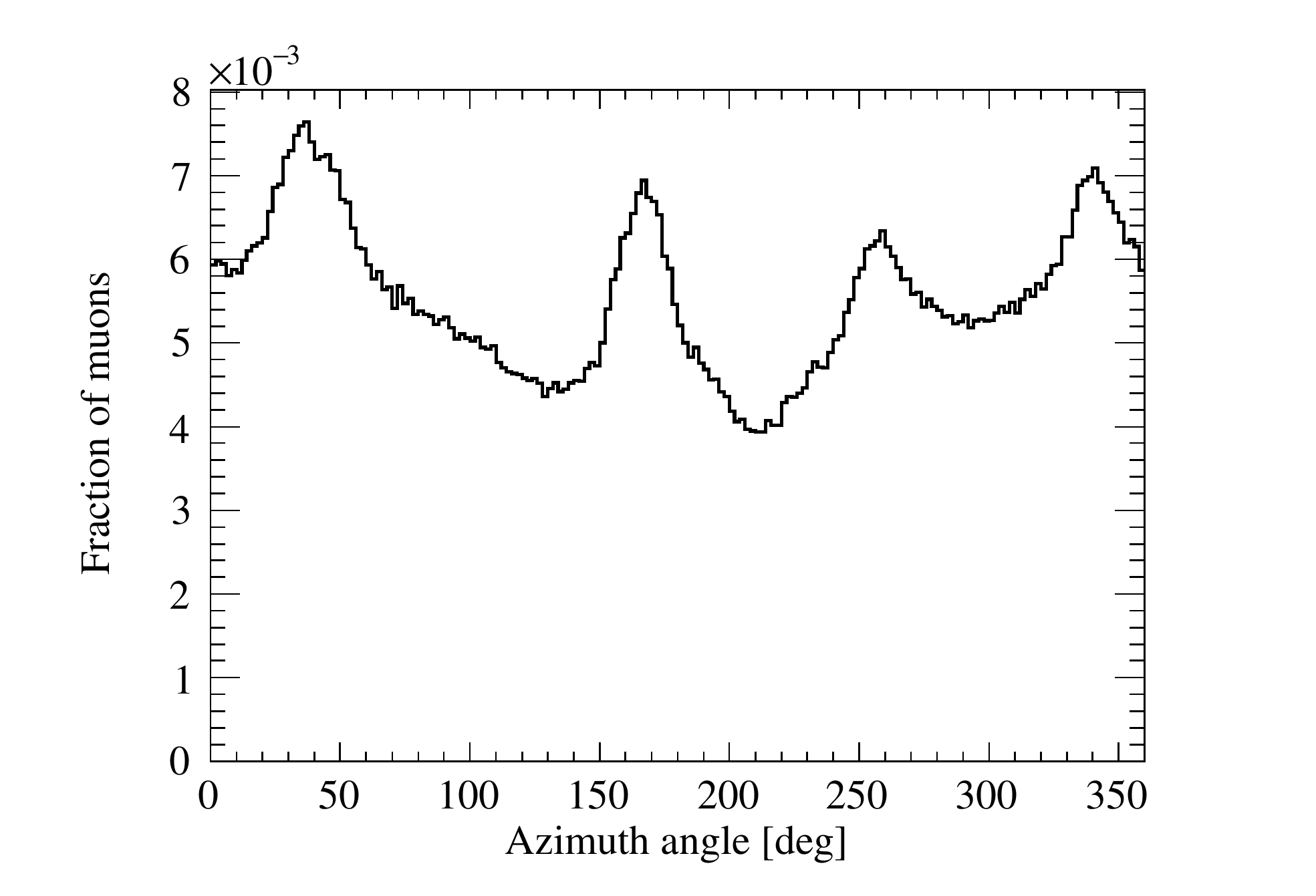}
\caption {Top: surface profile around SURF. The position of the LZ detector is shown by the black circle in the middle. The color scheme depicts the altitude above sea level in meters. East direction is to the right. Bottom: azimuthal angle distribution of $10^7$ muons at SURF as generated by MUSUN; azimuth angle is counted from East to North. Muon intensity is integrated over zenith angle.}
\label{fig:muons}
\end{center}
\end{figure}

Figure \ref{fig:muons} shows the surface profile around SURF (top) and the azimuthal angular distribution of muons at SURF integrated over zenith angle (bottom). The open cut in the surface profile (shown as a blue region on the top plot) results in a peak in muon intensity at about 170 degrees counted from East to North.

\subsubsection{Muon-induced neutrons}
\label{sec:muonneutrons}
Recorded energy spectra and angular distributions of muons were used to generate muon events within LUXSim \cite{luxsim:2012} (a predecessor of BACCARAT with a similar performance; BACCARAT was not available at the time of these simulations but the physics was the same in both codes). Muons were sampled on the surface of a box that encompasses the cavern and a few meters of rock around it (7~m on top and 5~m from all other sides), to account for muon-induced cascades that can start in rock and propagate to the detector. Rock composition and density were taken from \cite{rock} and the detector geometry and physics processes were similar to those in the current version of BACCARAT. 

\begin{figure}[htb]
\begin{center}
\includegraphics[width=8cm, trim=1cm 0 1cm 0, clip]{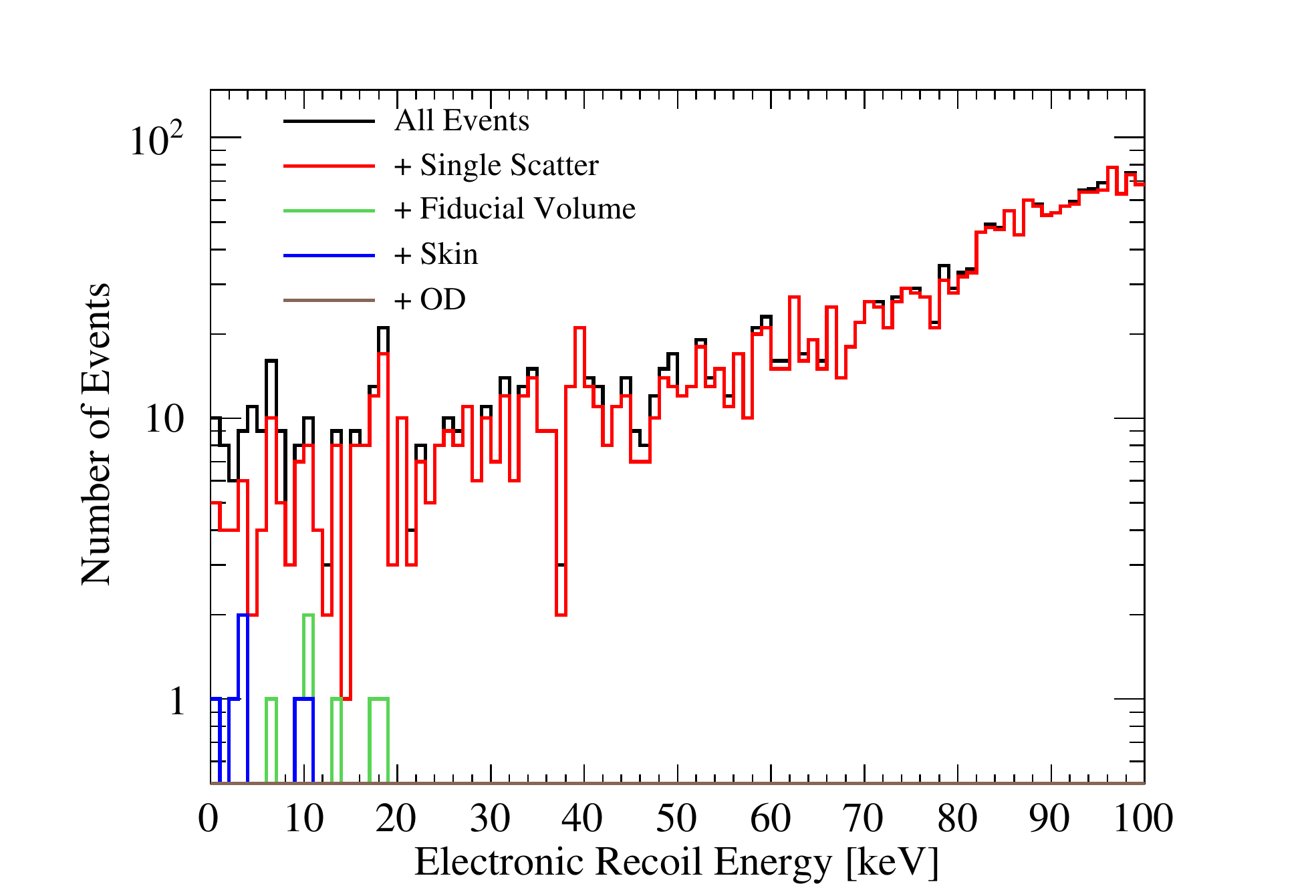}
\caption {Energy spectra of all muon-induced events in the liquid xenon TPC. The spectra are shown for events surviving each of the analysis cuts applied consecutively. $2.3\times10^{8}$ muons corresponding to approximately 120 years of live time were simulated in LUXSim and analysed applying standard background rejection cuts. No event survived all cuts in the 6 -- 30 keV$_{nr}$ range.}
\label{fig:muon-neutrons}
\end{center}
\end{figure}

In total, $2.3\times10^{8}$ muons corresponding to $\approx$120 live years were simulated and analyzed applying the background rejection cuts described in Section \ref{sec:overview}. In the 6--30 keV$_{nr}$ energy range, there are $1.4\pm0.2$ `pure' NR events (i.e. with no other energy deposition) in 1000 days before any event selections are made. The energy spectra of all muon-induced events surviving each of the cuts applied consecutively are shown in Figure \ref{fig:muon-neutrons}. Most events at these energies are single scatters that occur outside the fiducial volume. 

Of the small number of events that remain, all are removed by the LZ veto systems (the skin and outer detector). In this analysis the water tank, which has a high probability to veto events by detecting Cherenkov light, has not been considered. In any case, there are no events surviving all cuts, which allows a limit to be set on the number of background events arising from muon-induced sources. Using a Feldman-Cousins approach \cite{F-C}, the upper limit is 0.056 muon events at 90\% confidence level in 1000 live days and 5.6 tonnes of fiducial mass of LXe. This rate is sub-dominant to nuclear recoil backgrounds from the PTFE walls of the TPC; LZ detector components such as the cryostat vessels, PMTs and TPC assemblies; and also from atmospheric and diffuse supernova neutrinos. 

\subsection{Ion Emission from PTFE Walls}
\label{sec:wall}
\begin{figure}[htb]
\begin{center}
\includegraphics[width=8cm, trim=1cm 0 1cm 0, clip]{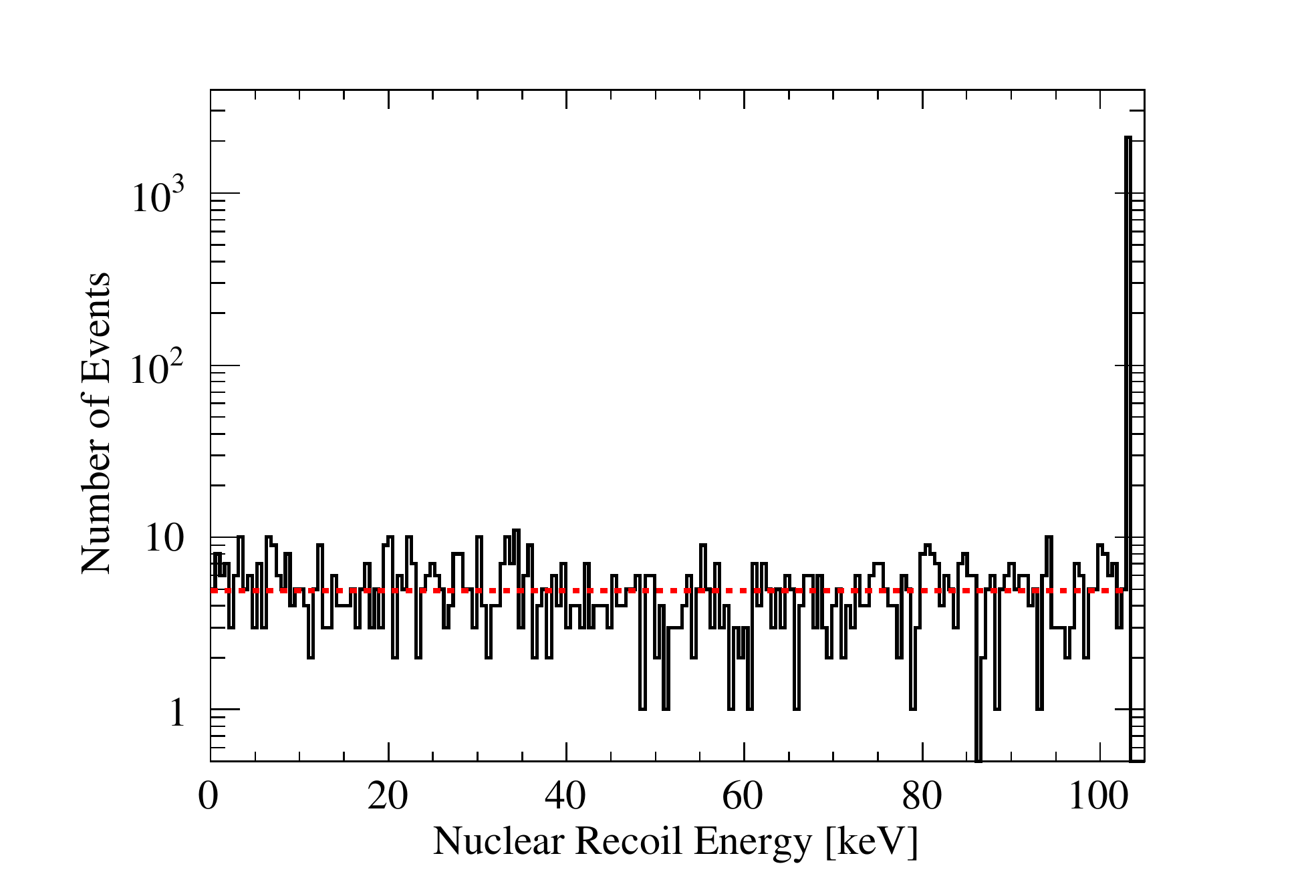}
\caption{Spectrum of energy deposits (solid) in LXe from 10,000 recoiling $^{206}$Pb nuclei following $^{210}$Po decays on and close to the surface of PTFE. Approximately 67\% of $^{206}$Pb recoils start on the PTFE surface resulting in a 103~keV peak whereas the remaining recoils have a kinetic energy less than 103~keV (flat part of the spectrum) due to $^{210}$Po atoms buried in the PTFE by some tens of nanometres. The fit (dashed line) to this spectrum and the peak are used as inputs to the wall event generator. In practice, the rate of 103~keV events may be reduced by cleaning $^{210}$Po off the surface of the PTFE.}
\label{fig:wall-events}
\end{center}
\end{figure}

The final stable product in the $^{222}$Rn decay chain is $^{206}$Pb, which is emitted with 103 keV kinetic energy during the $\alpha$-decay of its parent $^{210}$Po.  This decay may result in a continuum spectrum of nuclear-recoil signals up to around 103~keV depending on the implantation depth of $^{210}$Po into surfaces (e.g. PTFE walls) in contact with the active LXe volume. The contamination of PTFE by $^{210}$Po usually comes from its exposure to air in a clean room. We are using an extension~\cite{LRT2017MorrisonPlateout} of the so-called `Jacobi' model \cite{Jacobi1958,knutsonJacobi} applied to typical clean rooms to predict the plate-out of radon progeny onto PTFE surfaces. SRIM data \cite{Ziegler2010} were used to predict the migration of the radon progeny due to subsequent decays within, into, and out of the PTFE. In our toy model, we consider all radon and its progeny decaying consecutively. Of all $^{210}$Po decays, roughly 30\% produce $^{206}$Pb recoils that will enter the liquid xenon. Of these, about 67\%  have kinetic energy of 103\,keV, due to the $^{210}$Po atom starting on the PTFE surface, and the rest $<103$\,keV, with the $^{210}$Po located tens of nanometres within.  The spectrum of energy deposits is illustrated in Figure \ref{fig:wall-events}. The background from wall events will be rejected with high efficiency with position cuts.

\section{Simulations of Calibration Sources}
\label{sec:Calibrations}

LZ will be calibrated with a suite of sources producing both electron and nuclear recoils in the TPC, skin, and outer detectors. Sources can be deployed as `internal' sources, which are dissolved in the liquid xenon; `external' sources, which are encapsulated and introduced via one of three source tubes; or beam sources that fire energetic particles from outside the detectors. Table~\ref{tab:sources} contains a list of sources that will be used for LZ calibration.

\begin{table}
\caption{\label{tab:sources}LZ calibration sources}
\begin{ruledtabular}
\begin{tabular}{lll}
  Source & Deployment & Type\\
\hline
  CH$_3$T & Internal & $\beta$ \\
  $^{131m}$Xe & Internal & $\gamma$\\
  $^{83m}$Kr & Internal & $\beta$/$\gamma$\\
  $^{220}$Rn & Internal & $\alpha$/$\beta$ \\
  \hline
  $^{22}$Na & External & $\beta^+$/$\gamma$\\
  $^{57}$Co & External & $\gamma$\\
  $^{228}$Th & External & $\gamma$\\
  AmLi & External & neutron\\
  $^{88}$YBe & External & neutron\\
  DD & Beam & neutron\\
\end{tabular}
\end{ruledtabular}
\end{table}

Of these sources, $^{22}$Na and $^{57}$Co are simple nuclear decay sources that are generated according to \geantFour internal mechanisms, and $^{220}$Rn and $^{228}$Th are part of the $^{232}$Th decay chain, where the implementation is described in Section~\ref{sec:Backgrounds}. The other sources have custom implementations for LZ.

\textbf{CH$_3$T}: Tritiated methane can be dissolved in liquid xenon to produce a uniformly distributed source of electron recoil events. This source spans a range of energies from 0 to the tritium endpoint of~18.6 keV, and the methane can easily be extracted from the xenon when the calibration is complete. This source was used to great effect in LUX~\cite{akerib:2015wdi}.

The default behavior in \geantFour is that tritium is stable, so the decay spectrum for tritium is implemented by hand according to the standard beta decay spectrum. The end point used comes from Ref.~\cite{nagy:2006q}.

\textbf{$^{83m}$Kr}: $^{83m}$Kr is another source that can be distributed uniformly through the detector and used to calibrate position response ~\cite{Kr83m}. It has an additional benefit of two consecutive decays that can be used to calibrate recombination response and electric field. 

This decay occurs in two steps, separated by a short decay half-life ($\tau=154$~ns): the first transition of 32.1~keV and the second transition of 9.4~keV, where both steps can emit internal conversion electrons with or without Auger electrons, or a single $\gamma$. 

LZ generator implements the relative branching fractions for these steps according to the calculations from Ref.~\cite{aprile:2012an} and the time separation between the two stages of decay. This is critical for handling the electron recombination effects in liquid xenon.

\textbf{$^{131m}$Xe}:  Due to the longer mixing time scales required for the 10~t of liquid xenon to be used in LZ, and the relatively short half-life for $^{83m}$Kr (1.8 hours), the spatial distribution of krypton may not become fully homogeneous before completely decaying, which may reduce its usefulness for position calibration. For this reason $^{131m}$Xe, with a half life of 11.9 days, has been chosen as an additional calibration source.

Like $^{83m}$Kr, $^{131m}$Xe can emit internal conversion electrons or Auger electrons which are not modeled well by in-built \geantFour processes. Based on the work reported in Ref.~\cite{Ringbom}, correct distributions of gamma-rays, conversion and Auger electrons are implemented manually.

\begin{figure}
\centering
\includegraphics[width=\linewidth]{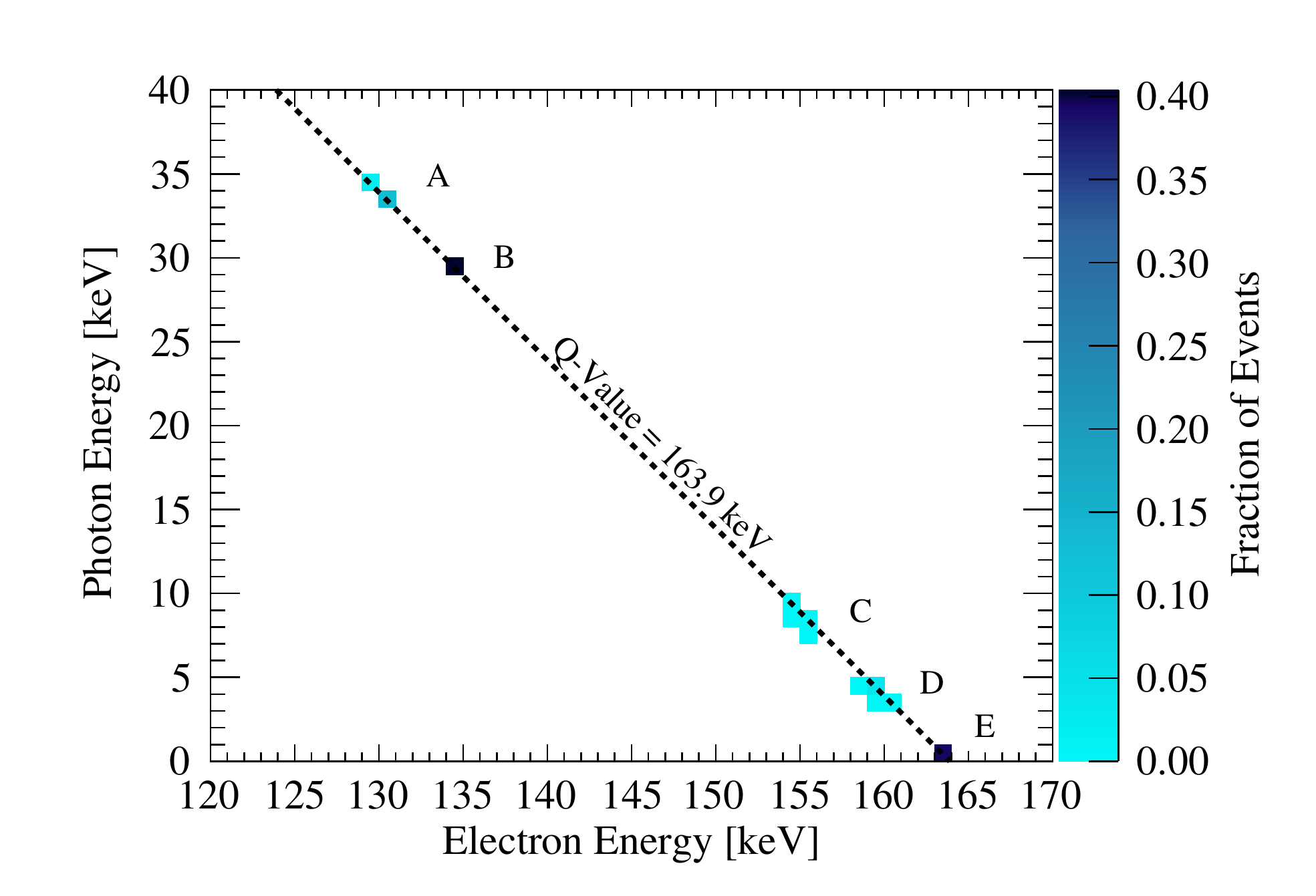}
\caption{\label{fig:xe131m} Photon energy versus electron energy for $^{131m}$Xe decays. The sum of the two always gives 163.93\,keV total energy deposition. The colour scheme shows the fraction of events in a particular region. Labels A -- E show different populations formed by various combinations of conversion electrons, Auger electrons and X-rays.}
\end{figure}

The energy difference between the metastable state and the ground state of $^{131}$Xe is 163.93~keV and a single gamma-ray of this energy is emitted 1.95\% of the time. The remaining cases consist of a combination of conversion electrons, Auger electrons and X-rays from atomic relaxation. In each case decay products are released in conjunction to add up to 163.93~keV. Figure \ref{fig:xe131m} shows the resulting scatter plot of energy deposition from photons against that from electrons arising from this isomeric state transition. 

The metastable state $^{131m}$Xe is produced in the decay of $^{131}$I (0.39\% of $^{131}$I decay to the metastable state of Xe) available in medical industry. 

\begin{figure}
\centering
\includegraphics[width=\linewidth]{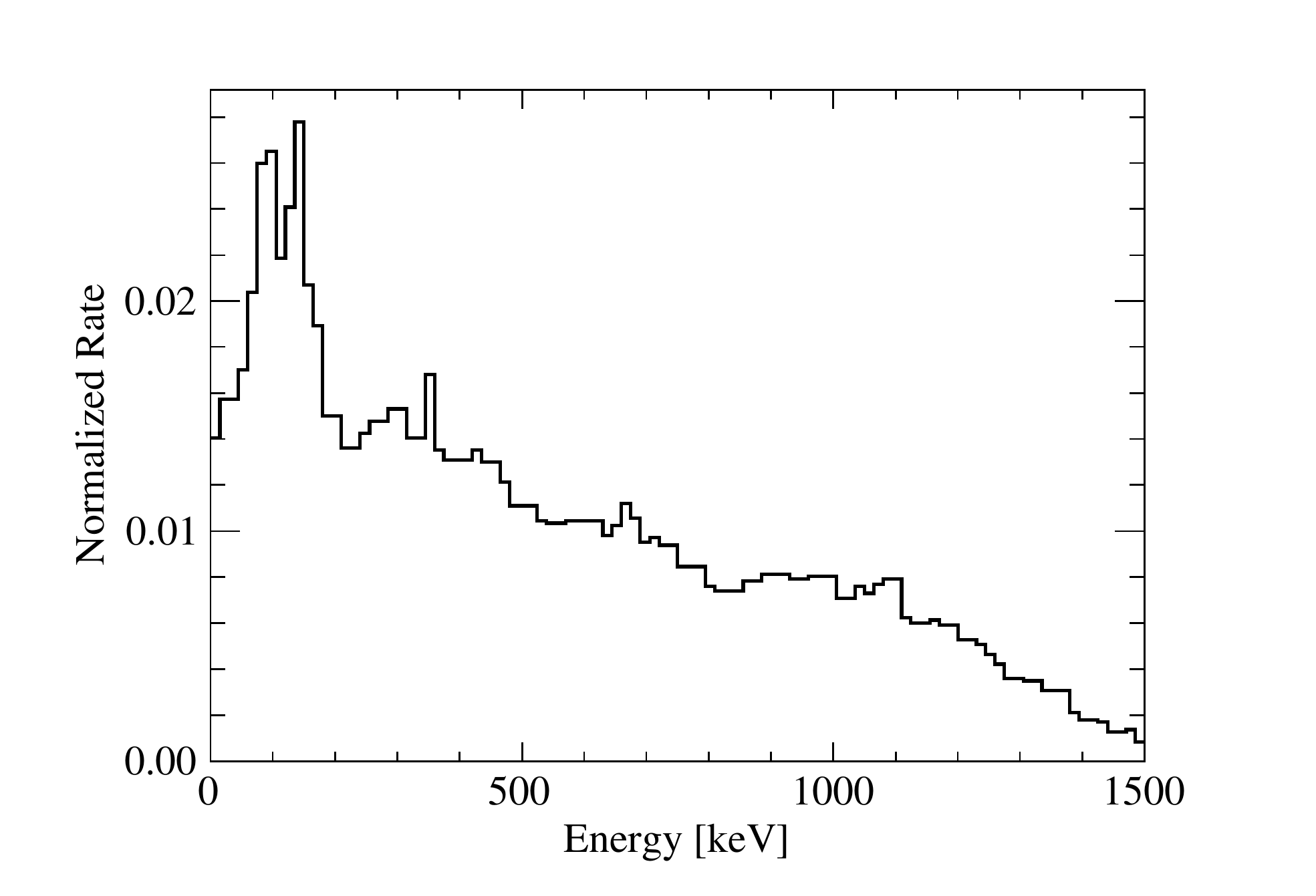}
\caption{\label{fig:AmLi} Measured neutron energy spectrum from a prototype AmLi source for the LZ experiment.}
\end{figure}

\textbf{AmLi}: AmLi is a desirable neutron source for dark matter calibration because the endpoint of the $\alpha$ + $^7\mbox{Li}$ $\rightarrow$ $^{10}\mbox{B}$ + $n$  reaction is relatively low (about 1.5~MeV), producing nuclear recoils predominantly in the dark matter search region. The LZ AmLi source is a custom source built at the University of Alabama. During fabrication, a prototype neutron spectrum (Figure~\ref{fig:AmLi}) was measured with $^3$He proportional counters. The simulated source is implemented from the measured spectrum of the prototype source, and will be updated to the measured spectrum of the production source when it is ready. 

\begin{figure}
\centering
\includegraphics[width=\linewidth]{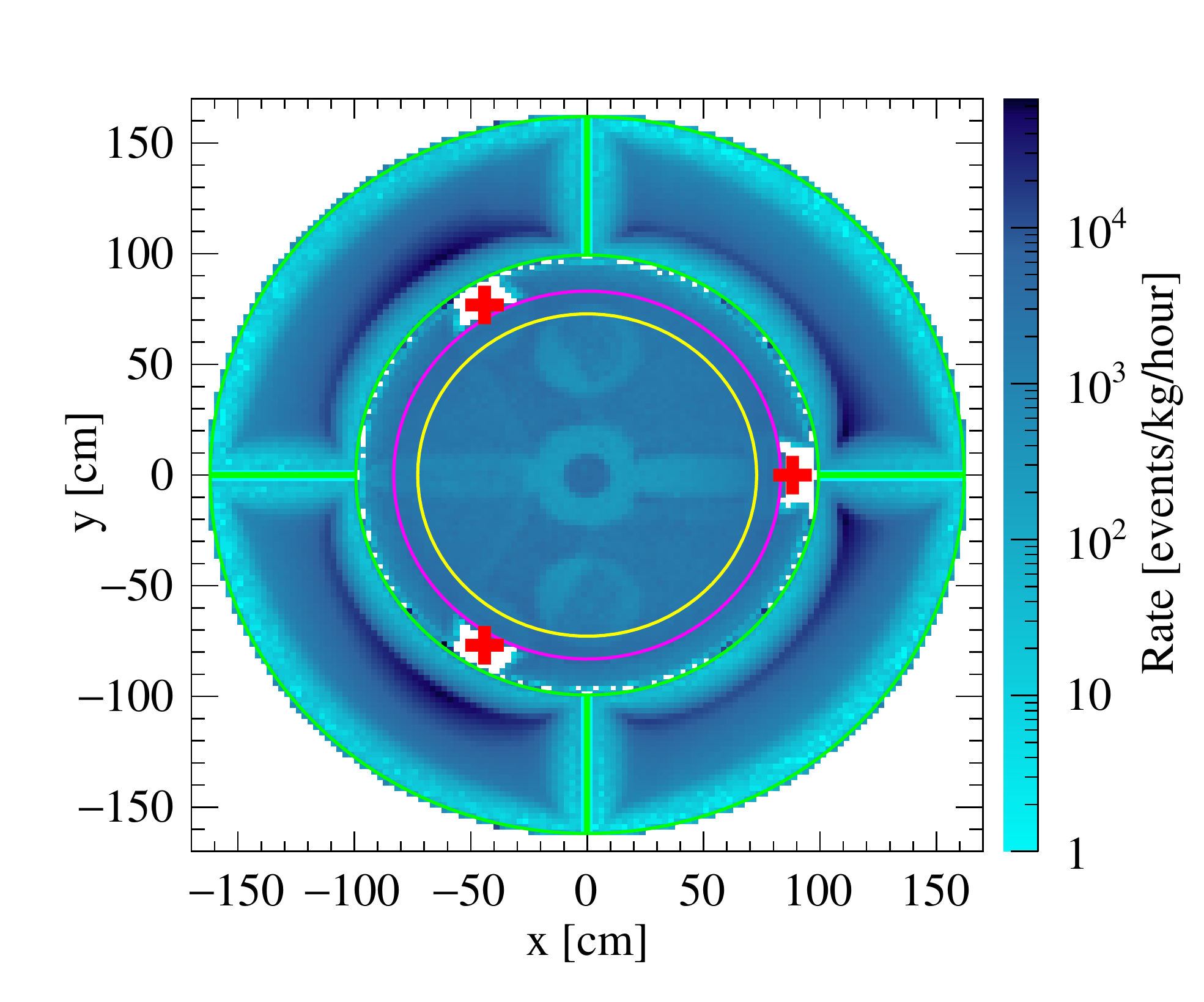}
\caption{\label{fig:AmLiT} Locations of AmLi NRs in the xenon and GdLS in the XY plane, with a source placed at the mid-height of the TPC in each of the three source tubes (red crosses). The TPC is marked in yellow, the skin in pink and the side tanks of the outer detector in green. Substructure of the bottom outer detector tank under the TPC can be seen in the yellow-bounded region.}
\end{figure}

Figure \ref{fig:AmLiT} shows positions of NRs in the liquid xenon and organic scintillator caused by irradiation of the detector with neutrons from AmLi sources, assuming an intensity of 130\,n/s. With three sources deployed at different positions inside three tubes, the whole TPC can be irradiated with neutrons.

\textbf{$^{88}$YBe}: $^{88}$YBe is a photoneutron source that produces low energy, almost monoenergetic (153 keV) neutrons through the process $\gamma$ + $^9\mbox{Be}$  $\rightarrow$ $n$ + $^8\mbox{Be}$ ~\cite{collar:2013xva}. A very strong $^{88}$Y source is needed to give a sufficient quantity of neutrons for a viable photoneutron source as the $(\gamma, n)$ cross section is quite small. For practical purposes, a large amount of tungsten shielding reduces the gamma rate in the detector while preserving a useful neutron flux. The current generator in BACCARAT has two production modes, one which generates gammas directly from $^{88}$Y decays in \geantFour and another which samples isotropic vectors to produce photoneutrons with appropriate position, energy, and momentum distributions. These generator modes can be sampled with the appropriate ratio according to the $(\gamma, n)$ cross section, or they can be sampled independently and scaled as needed to more efficiently produce photoneutron statistics.

\textbf{DD}: The DD neutron generator produces 2.45 MeV monoenergetic neutrons. The DD source was a critical calibration for LUX~\cite{akerib:2016mzi}, providing a source of neutron scatters with precisely known energy. In LZ, these neutrons reach the TPC through one of two tubes that penetrate the outer detectors. The simulation of this source is implemented very simply: as a fixed-position monoenergetic source of neutrons, which are fired within a small solid angle along the direction of the tubes. 


\section{Simulations of Rare and Forbidden Decays of Xe Isotopes}
$^{136}$Xe and $^{134}$Xe are two naturally occurring isotopes of xenon that are theoretically predicted to undergo two-neutrino double beta decay ($2\nu \beta \beta$), a Standard Model process involving the emission of two electrons and two neutrinos. Of these, only the $^{136}$Xe decay has been observed, with a measured half-life of $2.165 \times 10^{21}$ years \cite{exo:2012}. For the full LZ exposure, a total of 67 ER events are expected in the 1.5-6.5 keVnr WIMP search ROI from $^{136}$Xe decay \cite{LZ-sensitivity}, and with a Q-value of 2.458 MeV this process will also present a non-negligible background in a number of rare ER event searches, such as neutrinoless double beta decay ($0\nu \beta \beta$). $0\nu \beta \beta$ is an alternative mechanism through which these isotopes may decay, where two electrons are emitted with a combined energy equal to the Q-value of the decay. Due to the low-background detector environment, LZ will be able to search for $0\nu \beta \beta$ of $^{136}$Xe.

\begin{figure}[!h]
\centering
\includegraphics[width=.45\textwidth]{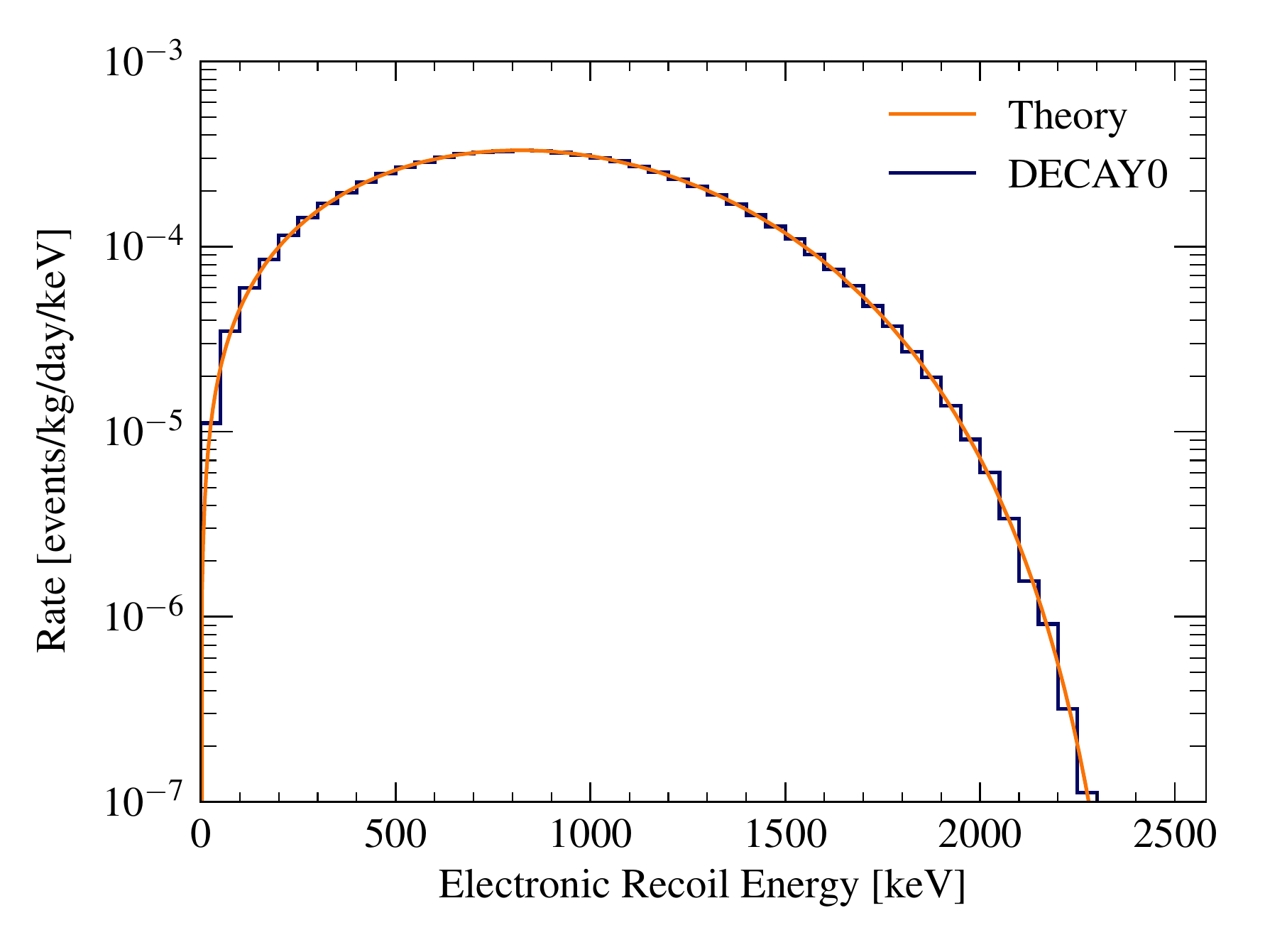}
\caption{Summed electron recoil energy spectrum of the $^{136}$Xe $2\nu \beta \beta$ decay. Theoretical curve (orange) from \cite{Kotila} is superimposed on the simulated histogram (black) from the DECAY0 generator \cite{Ponkratenko:2000um}.} 
\label{figure:Xe136_2nBB}
\end{figure}

In order to understand the detector response and quantify the detection efficiency for $0\nu \beta \beta$ and $2\nu \beta \beta$ processes, the DECAY0 generator \cite{Ponkratenko:2000um} was interfaced with BACCARAT to generate the initial kinematics of the emitted electrons. The $^{136}$Xe $2\nu \beta \beta$ spectrum from \cite{Kotila} (Figure~\ref{figure:Xe136_2nBB}), which was used to characterise this background in $^{136}$Xe $0\nu \beta \beta$ search ROIs for the LZ sensitivity estimates \cite{LZ-sensitivity}, agrees well with that obtained using the DECAY0 generator.

$^{124}$Xe and $^{126}$Xe are also present in natural xenon and are also expected to decay, but being on the proton-rich side of the even-even isobars mass parabola they will undergo double electron capture ($2\text{EC}$). Additionally, the high Q-value of $^{124}$Xe (2.864 MeV) opens up the possibility for two additional decay modes: $2\nu \beta^+ \beta^+$ and $2\nu \beta^+ \text{EC}$. Of these modes only $2\nu\,2\text{EC}$ has been observed in xenon, with a half-life of $1.8 \times 10^{22}$ years \cite{Xe124_2EC}. To study these decays in LZ, generators were developed for ${}^{124}$Xe $2\nu\,2\text{EC}$ and  $2\nu \beta^+ \text{EC}$ (the half-life for $2\nu \beta^+ \beta^+$ is expected to be too long to be observable in LZ). In the former only the atomic de-excitation can be detected, and this is simulated by generating two isotropic X-rays each with half of the visible energy. For $2\nu \beta^+ \text{EC}$, the identifying feature of the decay will be the unique topology of the positron signal in the detector, and the energy of the positron will dominate the atomic de-excitation. Therefore, the $2\nu \beta^+ \text{EC}$ simulation is solely of the positron, with it emitted isotropically with energy following the spectrum obtained from the Beta Spectrum Generator \cite{Hayen_2019}, eventually annihilating and producing two 511\,keV gamma-rays.


\section{Simulations of WIMPs and Neutrinos}

\subsection{WIMPs}
\label{sec:WIMPs}
The generation of NR events from WIMP spin-independent interactions for sensitivity studies uses the `standard' isothermal halo model as given in \cite{McCabe} and the scattering formalism in \cite{lewin:1995rx}. Given the expected spatial uniformity and the clean, 'golden,' single S1-S2 topology of the events, a signal PDF could be produced for given WIMP masses in LZ by passing a parameterization of the detector response and the derived recoil energies to NEST for corrected S1 and S2 calculations under its NR model. Changes to the halo model parameters and nuclear physics (for example, in considering spin-dependent interactions or non-standard WIMP scenarios) can therefore be accommodated by propagating the effects through to the underlying recoil spectrum.

For the end goal of simulating waveforms, that same recoil spectrum can be first sampled by the spectrum generator in BACCARAT. The spectrum generator takes as input a detector volume, spectrum, and type of particle to generate events. For WIMPs, it can be used to produce recoiling xenon nuclei, and the resultant energy depositions can be processed much as described before for the MDC chain and other generators (Section~\ref{sec:overview}). The ability to generate individual atoms allows us to take into account the relative isotopic abundances, as would be important in the case of spin-dependent studies.

\subsection{Astrophysical and Atmospheric Neutrinos}
\label{sec:Neutrinos}
Electron and nuclear recoils will be produced in LZ by solar neutrinos, diffuse supernova neutrinos and atmospheric neutrinos. Coherent elastic neutrino-nucleus scattering will be responsible for NR events. In calculating the LZ sensitivity to WIMPs \cite{LZ-sensitivity}, neutrino fluxes and spectra for solar neutrinos from \cite{bahcall:2004mz} and oscillation parameters from \cite{Olive:2016xmw} were used. As outlined for WIMPs (Section~\ref{sec:WIMPs}), this information was used directly in NEST for signal generation for this application, but the spectra can also be used as input to the spectrum generator to simulate waveforms from associated NR and ER events.

Solar neutrinos do not contribute to the NR background in LZ for a high-mass WIMP search ($>20$\,GeV/$c^2$). However, atmospheric neutrinos produce NRs at higher energies and this background constitutes one of the largest contributions to the total NR background in LZ.


\section{Conclusions}
\label{sec:Conclusions}
The LZ detector is projected to explore new parameter space in the search for WIMPs, as well as other dark matter candidates and various other rare physics phenomena. Monte Carlo simulations, which encapsulate signal, background and calibration sources and their effects inside of the LZ detector, are essential to performing these searches.  

The simulations described here utilize a \geantFour-based package called BACCARAT (or its predecessor LUXSim for muon-induced neutrons). A suite of custom event generators and physics models have been integrated to accurately describe particle generation, propagation and the subsequent interaction inside of the LZ detectors.

These features enable LZ to generate a background model, which has so far been used to assess the sensitivity for detecting WIMPs ~\cite{LZ-sensitivity}. The specific background contributions of various sources have been studied and compared, with radioactivity from radon dispersed within the liquid xenon the largest contributor to the total background rate.           

Additional software in the form of a Detector Electronics Response (DER) code is capable of producing mock waveform data using the BACCARAT output. The event reconstruction software (LZap, not covered here) has been developed using this mock data, and ensures the experiment can be ready for physics analyses as soon as data-taking begins in 2020.


\section{Acknowledgements}
The research supporting this work took place in whole or in part at the Sanford Underground Research Facility (SURF) in Lead, South Dakota. Funding for this work is supported by the U.S. Department of Energy, Office of Science, Office of High Energy Physics under Contract Numbers DE-AC02-05CH11231, DE-SC0020216, DE-SC0012704, DE-SC0010010, DE-AC02-07CH11359, DE-SC0012161, DE-SC0014223, DE-FG02-13ER42020, DE-SC0009999, DE-NA0003180, DE-SC0011702,  DE-SC0010072, DE-SC0015708, DE-SC0006605, DE-FG02-10ER46709, UW PRJ82AJ, DE-SC0013542, DE-AC02-76SF00515, DE-SC0019066, DE-AC52-07NA27344 and DOE-SC0012447. This research was also supported by U.S. National Science Foundation (NSF); the U.K. Science \& Technology Facilities Council under award numbers, ST/M003655/1, ST/M003981/1, ST/M003744/1, ST/M003639/1, ST/M003604/1, and ST/M003469/1; Portuguese Foundation for Science and Technology (FCT) under award numbers PTDC/FIS-PAR/28567/2017; the Institute for Basic Science, Korea (budget numbers IBS-R016-D1); University College London and Lawrence Berkeley National Laboratory thank the U.K. Royal Society for travel funds under the International Exchange Scheme (IE141517).

We acknowledge additional support from the STFC Boulby Underground Laboratory in the U.K., the GridPP Collaboration \cite{GridPP1} \cite{GridPP2}, in particular at Imperial College London and additional support by the University College London (UCL) Cosmoparticle Initiative. This research used resources of the National Energy Research Scientific Computing Center, a DOE Office of Science User Facility supported by the Office of Science of the U.S. Department of Energy under Contract No. DE-AC02-05CH11231. The University of Edinburgh is a charitable body, registered in Scotland, with the registration number SC005336. The assistance of SURF and its personnel in providing physical access and general logistical and technical support is acknowledged. \newline



\pagebreak
\onecolumngrid
\vspace{1cm}
\appendix
\section{\bf Neutron Yields}
\label{app:neutron-yields}

\begin{table}[htb]
\centering
\caption{Neutron yield from ($\alpha,n$) reactions in different materials. The column ``Abundance" gives the chemical composition of the source used to calculate neutron spectra with the abundance of elements (by the number of atoms, not mass) given in brackets. Only elements with the abundance greater than 1\% are shown (with the accuracy of 1\%). Neutron yield (columns 3--6) is shown as the number of neutrons per gram of material per second per ppb of U and Th concentration. Uranium and thorium decay chains are assumed to be in equilibrium in columns 3 and 6. In columns 4 and 5 early and late uranium sub-chains are shown separately. $^{235}$U is added to the early sub-chain. Spontaneous fission is significant for $^{238}$U only and is independent of the material with a neutron yield of $1.353 \times 10^{-11}$~n/g/s/ppb.}
\vspace{0.2cm}
\begin{tabular}{|c|c|c|c|c|c|}
\hline
  \multicolumn{2}{|c|}{} & \multicolumn{4}{|c|}{Neutron yield in n/g/s/ppb}  \\ \hline
    Material &   Abundance, \%                   &   U &  U$_{early}$ & U$_{late}$ & Th \\
\hline
PTFE   & C(33),F(67)   & $8.72\times10^{-10}$ & $1.36\times10^{-10}$ & $7.36\times10^{-10}$ & $3.50\times10^{-10}$ \\
\hline
Aluminum   & Al(100)   & $1.69\times10^{-10}$ & $1.46\times10^{-11}$ & $1.54\times10^{-10}$ & $8.59\times10^{-11}$\\
\hline
Ceramics   &  Al(40),O(60)   & $8.59\times10^{-11}$ & $7.76\times10^{-12}$ & $7.81\times10^{-11}$ & $4.32\times10^{-11}$ \\
\hline
Copper   &  Cu(100)   & $3.11\times10^{-13}$ & $8.42\times10^{-15}$ & $3.03\times10^{-13}$ & $9.70\times10^{-13}$ \\
\hline
Titanium   &  Ti(100)   & $2.55\times10^{-11}$ & $1.11\times10^{-12}$ & $2.44\times10^{-11}$ & $2.15\times10^{-11}$ \\
\hline
Acrylic   &  C(13),O(33),H(54)  & $1.30\times10^{-11}$ & $2.33\times10^{-12}$ & $1.07\times10^{-11}$ & $5.05\times10^{-12}$ \\
\hline
Stainless steel   &  Fe(66),Ni(12),Cr(17),Mn(2),Mo(3)  & $4.93\times10^{-12}$ & $1.85\times10^{-13}$ & $4.76\times10^{-12}$ & $5.77\times10^{-12}$ \\
\hline
Quartz   &  Si(33),O(67)  & $1.59\times10^{-11}$ & $2.01\times10^{-12}$ & $1.39\times10^{-11}$ & $7.02\times10^{-12}$ \\
\hline
Polyethylene   &  C(33),H(67)  & $1.43\times10^{-11}$ & $2.56\times10^{-12}$ & $1.18\times10^{-11}$ & $5.61\times10^{-12}$ \\

\hline                         
\end{tabular}
\label{table-neutrons}
\end{table}

\end{document}